\pdfoutput=1

\documentclass[a4paper,fleqn,usenatbib]{mnras}

\usepackage{mathptmx}

\usepackage[T1]{fontenc}
\usepackage{ae,aecompl}

\usepackage{graphicx}	
\usepackage{amsmath}	
\usepackage{amssymb}	
\usepackage[pdftex]{thumbpdf}
\usepackage{lmodern}
\usepackage{natbib}
\usepackage{longtable}
\usepackage{pdflscape}
\usepackage[caption=false]{subfig}
\usepackage{epstopdf}
\usepackage{multicol}        
\usepackage{bm}		
\usepackage{pdflscape}	

\title[Large-scale structures in WR 134]{An extensive spectroscopic time-series of three Wolf-Rayet stars. I. The lifetime of large-scale structures in the wind of WR 134}

\author[E. J. Aldoretta et al.]{E. J. Aldoretta,$^{1}$
N. St-Louis,$^{1}$
N. D. Richardson,$^{2}$
A. F. J. Moffat,$^{1}$ \newauthor
T. Eversberg,$^{3,4,5}$
G. M. Hill,$^{6}$
T. Shenar,$^{7}$ 
\'E. Artigau,$^{1}$
B. Gauza,$^{8,9}$ \newauthor
J. H. Knapen,$^{8,9}$ 
J. Kub\'at,$^{10}$ 
B. Kub\'atov\'a,$^{10,11}$ 
R. Maltais-Tariant,$^{1}$ 
M. Mu\~{n}oz,$^{1}$ \newauthor
H. Pablo,$^{1}$ 
T. Ramiaramanantsoa,$^{1}$ 
A. Richard-Laferri\`ere,$^{1}$ 
D. P. Sablowski,$^{4,5,12}$ \newauthor
S. Sim\'on-D\'iaz,$^{8,9}$  
L. St-Jean,$^{1}$ 
F. Bolduan,$^{5}$ 
F. M. Dias,$^{4,5}$ 
P. Dubreuil,$^{5,13}$ \newauthor
D. Fuchs,$^{5}$ 
T. Garrel,$^{5,13}$ 
G. Grutzeck,$^{4,5}$ 
T. Hunger,$^{4,5}$ 
D. K\"usters,$^{4,5}$ \newauthor
M. Langenbrink,$^{5}$ 
R. Leadbeater,$^{4,5,14}$ 
D. Li,$^{4,5,15}$ 
A. Lopez,$^{5,13}$ 
B. Mauclaire,$^{5,13}$ \newauthor
T. Moldenhawer,$^{5}$ 
M. Potter,$^{5,16}$ 
E. M. dos Santos,$^{5}$ 
L. Schanne,$^{4,5}$ 
J. Schmidt,$^{5}$ \newauthor
H. Sieske,$^{5}$ 
J. Strachan,$^{5,17}$ 
E. Stinner,$^{5}$ 
P. Stinner,$^{4,5}$ 
B. Stober,$^{4,5}$ \newauthor
K. Strandbaek,$^{4,5}$ 
T. Syder,$^{5}$
D. Verilhac,$^{5,13}$ 
U. Waldschl\"ager,$^{4,5}$
D. Weiss,$^{4,5}$ \newauthor
A. Wendt$^{5}$
\\
$^{1}$D\'epartement de Physique, Universit\'e de Montr\'eal, C.P. 6128 Succ. Centre-ville, Montr\'eal, QC, H3C 3J7, Canada\\
$^{2}$Ritter Observatory, Department of Physics and Astronomy, The University of Toledo, Toledo, OH 43606-3390, USA\\
$^{3}$Schn\"orringen Telescope Science Institute, Waldbr\"ol, Germany\\
$^{4}$VdS Section Spectroscopy, Germany (http://spektroskopie.fg-vds.de/index\_e.htm)\\
$^{5}$Teide Pro-Am Collaboration\\
$^{6}$W. M. Keck Observatory, 65-1120 Mamalahoa Highway, Kamuela, HI 96743, USA\\
$^{7}$Institut f\"ur Physik und Astronomie, Universit\"at Potsdam, Karl-Liebknecht-Str. 24/25, D-14476 Potsdam, Germany \\
$^{8}$Instituto de Astrof\'isica de Canarias, E-38200 La Laguna, Tenerife, Spain\\
$^{9}$Departamento de Astrof\'\i sica, Universidad de La Laguna, E-38206 La 
Laguna, Spain \\
$^{10}$Astronomick\'y \'ustav, Akademie v\v{e}d \v{C}esk\'e Republiky, 251 65 Ond\v{r}ejov, Czech Republic\\
$^{11}$Matemati\v{c}ki Institut SANU, Kneza Mihaila 36, 11001 Beograd, Serbia \\
$^{12}$Leibniz-Institut for Astrophysics Potsdam (AIP), An der Sternwarte 16, D-14482 Potsdam \\
$^{13}$Astronomical Ring for Access to Spectroscopy (ARAS), France (http://www.astrosurf.com/aras/)\\
$^{14}$Three Hills Observatory, The Birches, Torpenhow, Wigton, CA7 1JF, UK (http://www.threehillsobservatory.co.uk/astro/astro.htm) \\
$^{15}$Jade Observatory, 501-47-59, Jin Jiang Nan Li, Jin Jiang Road, He Bei District, Tianjin, China \\
$^{16}$Beverly Hills Observatory, P.O. Box 3626, Baltimore, MD 21214, USA (www.beverlyhillsastronomer.org) \\
$^{17}$School of Physics and Astronomy, Queen Mary University of London, 327 Mile End Rd., London, E1 4NS, UK 
}

\date{Accepted 2016 May 15. Received 2016 May 13; in original form 2016 March 11}

\pubyear{2016}

\begin{document}
\label{firstpage}
\pagerange{\pageref{firstpage}--\pageref{lastpage}}
\maketitle
\clearpage
\begin{abstract}
During the summer of 2013, a 4-month spectroscopic campaign took place to observe the variabilities in three Wolf-Rayet stars. The spectroscopic data have been analyzed for WR~134 (WN6b), to better understand its behaviour and long-term periodicity, which we interpret as arising from corotating interaction regions (CIRs) in the wind. By analyzing the variability of the He~{\sc II}~$\lambda$5411 emission line, the previously identified period was refined to $P =$ 2.255 $\pm$ 0.008 (s.d.) days. The coherency time of the variability, which we associate with the lifetime of the CIRs in the wind, was deduced to be 40 $\pm$ 6 days, or $\sim$ 18 cycles, by cross-correlating the variability patterns as a function of time. When comparing the phased observational grayscale difference images with theoretical grayscales previously calculated from models including CIRs in an optically thin stellar wind, we find that two CIRs were likely present. A separation in longitude of $\Delta \phi \simeq$ 90$^{\circ}$ was determined between the two CIRs and we suggest that the different maximum velocities that they reach indicate that they emerge from different latitudes. We have also been able to detect observational signatures of the CIRs in other spectral lines (C~{\sc IV}~$\lambda\lambda$5802,5812 and He~{\sc I}~$\lambda$5876). Furthermore, a DAC was found to be present simultaneously with the CIR signatures detected in the He~{\sc I}~$\lambda$5876 emission line which is consistent with the proposed geometry of the large-scale structures in the wind. Small-scale structures also show a presence in the wind, simultaneously with the larger scale structures, showing that they do in fact co-exist. 
\end{abstract}




\section{Introduction}

Massive stars dominate the ecology of the Universe, enriching the interstellar medium (ISM) through their dense stellar winds and terminal supernova explosions. They also contribute in a major way to the energetics of the ISM with their fast winds. Models have shown that mass-loss during the stellar lifetime will dominate the ISM enrichment (e.g., \citealt{Smith2006}) for stars with masses in excess $\sim$~60~$M_{\odot}$. An important aspect of these winds is their structure. On small scales, it has been known for some time that the winds are clumped \citep{Moffat1988}. Some massive stars have been proposed to also harbour large-scale structures that constitute a more global asymmetry. One promising model for this asymmetry is the development of corotating interaction regions (CIRs) in their winds as a result of a perturbation at their base \citep{Mullan1984}. The properties of such wind inhomogeneities have an enormous impact on derived mass-loss rates \citep{Hamann2008_workshop}, and thus on our understanding of stellar evolution \citep{Smith2014}. By studying the time-dependent behaviour of these structures, we can deepen our understanding of stellar winds. 

Amongst O stars, the most influential recent knowledge on stellar winds came through the study of UV resonance lines that exhibit P Cygni profiles \citep{Massa2015}. From an {\it International Ultraviolet Explorer} snapshot survey of about 200 stars, \citet{Howarth1989} found that most O stars show narrow absorption components (NACs) that are very likely snapshots of the time-variable discrete absorption components (DACs) seen in several intensive time-series of UV spectra of individual stars (e.g., \citealt{Massa1995a, Massa1995b, Howarth1995, Prinja1995,Kaper1996}). Further studies showed that there are typically two observable DACs per rotation cycle for massive stars \citep{Kaper1999}. These DACs can reflect the rotation cycle, but even when they do, they do not preserve coherency across many rotation cycles, meaning that the lack of detection of a strong global dipolar magnetic field was not entirely surprising for all well-studied DAC stars \citep{David-Uraz2014}. Recent results from optical photometry of a well-known star with DACs, $\xi$~Per (O7.5III(n)((f))), suggest that the star's photometric variability could be explained by bright surface spots that could drive the DAC behaviour \citep{Rami2014}. These spots could be the consequence of a small-scale (and therefore difficult to detect) localized magnetic field, arising from subsurface convection driven by an Fe opacity bump, as predicted by \citet{Cantiello2009} and \citet{Cantiello2011}. The same subsurface convection zone could also produce gravity waves that drive stochastic wind inhomogeneities (clumps) starting at the photosphere \citep{Cantiello2016}.

If CIRs were related to the internal structures and subsurface convection of massive stars, we may expect to find them in {\it evolved}, hot, massive stars as well. Wolf-Rayet (WR) stars are a particularly attractive test of this hypothesis, as their winds are some tenfold stronger than their H-burning O star progenitors, making any wind phenomena easier to detect in WR stars. These stars are mainly hydrogen deficient, helium-burning hot stars with strong winds ($\dot{M} \sim $ 10$^{-5}$ $M_{\odot}$ yrs$^{-1}$, v $\sim$ 1000 - 5000 km s$^{-1}$). However, the only well studied examples of WR stars that are suggested to harbour CIRs in their wind are WR~1, WR~6 and WR~134 \citep{Chene2010, Morel1997, Morel1999}. 

The strongest evidence for the presence of CIRs in a WR star wind is for WR~6 (EZ~CMa; \citealt{Robert1992, Georgiev1999, Lamontagne1986}), but it is in a difficult location for northern hemisphere observers. WR~134 (WN6), however, is bright ($V \sim$ 8.08) and located in the constellation Cygnus, making it an ideal target for large campaigns in the northern hemisphere to examine the CIR lifetime. \citet{Morel1999} studied the spectroscopic and photometric variability of WR~134, a massive star known to show variability with a period of $\sim$ 2.3 days \citep{McCandliss1994}. They confirmed the existence of a coherent 2.25 $\pm$ 0.05 day periodicity in changes of the He~{\sc II}~$\lambda$4686 emission line. Changes in other lines such as He~{\sc II}~$\lambda$5411 and He~{\sc II}~$\lambda$4542 were found to be correlated with the changes in the He~{\sc II}~$\lambda$4686 line. The global pattern of variability, however, was found to change with each epoch of observation. Interestingly, while this period was strongly detected within the spectroscopic data, it was only marginally detected in the photometric and broad-band polarimetric data. 

Previous analyses of these variable wind structures in WR stars have been interpreted as arising from the effects of a compact companion in orbit around the WR star (e.g., \citealt{Firmani1980}). However, these models and observations have improved, and the binary nature of several systems has been questioned in favour of CIRs (e.g., \citealt{Morel1998, Dessart2002}). \citet{Morel1999} performed simulated line-profile variations caused by the orbital revolution of a local and strongly ionized wind cavity from a collapsed companion. They were able to obtain a reasonable fit between the observed and modeled phase-dependent line profiles. However, the predicted X-ray flux from this cavity can only be reconciled with that observed if the accretion onto the collapsed companion is significantly inhibited, which would then not produce the cavity required to generate the line profile changes. Therefore, \citet{Morel1999} concluded that an interpretation based on rotational modulation of a large-scale structure in the wind offers greater consistency with the observations obtained. The line-profile changes in the He~{\sc II}~$\lambda$4686 line showed coherent, periodic patterns at all epochs reminiscent of the simulated line-profile variability from CIRs later published by \citet{Dessart2002}. The variations of the shape of the pattern can be attributed to the fact that these structures have an as yet undetermined finite lifetime. Throughout this paper, we follow this same approach and make the tactic assumption that the $\sim$periodic variability we find in the emission-line features of WR 134 reflects the rotational period in the same way, and that the large-scale features arise in CIRs.

The lifetime of these features actually constitutes a crucial question related to CIRs. \citet{Rami2014} had success in determining this for the mid-O giant, $\xi$ Per, with a month of {\it MOST} photometry, but similar observational efforts have not been performed for other massive stars, therefore leaving the proposed physical mechanism causing them, i.e. subsurface convection \citep{Cantiello2009}, unconstrained. In order to examine this for a WR star, we initiated a four-month long, worldwide professional/amateur spectroscopic campaign on WR~134 among three bright WR stars (the others being WR~135 and WR~137) during the summer of 2013, for which we present initial results in this publication. In Section 2, we present the spectroscopy of WR~134 that we collected from 10 sites across the globe. Section 3 discusses the measurements made for the isolated, strong He~{\sc II}~$\lambda$5411 emission line, along with a determination of the period of the variations to better precision than previous studies. Section 4 presents the results of the campaign, including the lifetime of the CIRs. We conclude this study with a discussion in Section 5. 

\section{OBSERVATIONS} 

The data collected for this project were taken between 25 May 2013 and 06 October 2013 with ten different telescopes. All data sets are tabulated in Table~\ref{tab:tab1}. The majority of the data were collected by the amateur astronomers among the co-authors (F.B., F.M.D., P.D., D.F., T.G., G.G., T.H., D.K., M.L., R.L., D.L., A.L., B.M., T.M., M.P., E.M.S., L.S., J.S., H.S., T.S., J.S., E.S., P.S., B.S., K.S., D.V., U.W., D.W., A.W.) at the Teide Observatory of the Instituto de Astrof\'isica de Canarias (IAC) in Tenerife using the 0.82m IAC80 telescope and a fiber-fed echelle spectrograph combined with a CCD, the latter two supplied by B.S. This eShel spectrograph was manufactured by Shelyak\footnote{http://www.shelyak.com/?lang=2}. Each night, a set of bias, dark and flat frames were collected and used in the calibration of the data. Wavelength calibration was accomplished with ThAr emission spectra obtained before or after each exposure. The reductions were accomplished through standard techniques with IRAF\footnote{IRAF is distributed by the National Optical Astronomy Observatory, which is operated by the Associated Universities for Research in Astronomy, Inc., under cooperative agreement with the National Science Foundation.}, but a significant number of images were found to be assembled incorrectly by the control computer. However, a significant dark current in the observations, including several brighter pixels in each row, allowed for the reassembly of these images through a custom-built row-by-row cross-correlation against a median-combined master dark that resulted in images that were indistinguishable from the standard spectral images. 

\begin{table*}
\begin{center}
\hspace*{-0.75cm}
\resizebox{\textwidth}{!}{
\begin{tabular}{lcccccccr}
\hline
\textbf{Observatory} & \textbf{Telescope} & \textbf{Spectrograph} & \textbf{CCD} & \textbf{HJD} & \textbf{$N_{spec}$} & \textbf{Resolving} & \textbf{$\lambda$ coverage} & \textbf{S/N} \\ 
  &  &  &  &\textbf{- 2,450,000} &  & \textbf{Power $\left({{\lambda}\over{\Delta\lambda}}\right)$} & \textbf{(\AA)} &  \\ [1ex]
\hline\hline
\multicolumn{9}{|c|}{{\it Professional Facilities}} \\
\hline
OMM (1) & 1.6m & Perkin-Elmer &  STA0520 Bleu & 6485 - 6496 & 19 & 5,300 & 4500 - 6700 & 350 \\
OMM (2) &      &            &               & 6561 - 6567 & 19 & 7,000 & 4750 - 6000 & 350 \\
DAO & 1.8m & Cassegrain &  SITe-2 & 6473 - 6482 & 58 & 5,500 & 5100 - 6000 & 350 \\
NOT & 2.5m & FIES & EEV 2k x 2k & 6468 & 1 & 11,800 & 3500 - 6800 & 350 \\
Ond\v{r}ejov & Perek (2m) & coud\'e & PyLoN 2048x512 BX & 6452 - 6549 & 21 & 10,000 & 5200 - 5800 & 300 \\
Keck & Keck II (10m) & ESI &  MIT-LL W62C2 & 6517 & 3 & 13,000 & 4000 - 10000 & 500 \\
Teide & IAC80 & eShel & Nova3200 & 6438 - 6550 & 184 & 10,500 & 4500 - 7400 & 250 \\
\hline
\multicolumn{9}{|c|}{{\it Amateur Contributors}} \\
\hline
Potter & C14 (0.36m) & LHIRES III & SBIG ST-8 & 6443 - 6572 & 36 & 7,500 & 5300 - 5900 & 200 \\
Li & C11 (0.28m) & LHIRES III & QHYIMG2P & 6489 - 6563 & 27 & 5,300 & 5200 - 5600 & 200 \\
Strachan & 0.25m & LHIRES III & ArtemisHSC & 6496 - 6505 & 17 & 5,300 & 5300 - 5700 & 200 \\
Leadbeater & C11 (0.28m) & LHIRES III & ATIK-314L+ & 6484 - 6507 & 10 & 5,300 & 5200 - 5600 &  200 \\ [0.5ex]
\hline
\end{tabular}}
\caption{List of each observatory that contributed to the campaign along with corresponding telescope, spectrograph, CCD information, the corresponding HJD interval at which the spectra were taken, the total number of spectra, the resolving power, the wavelength coverage and the average S/N ratio. The two rows for OMM represent the two separate runs, each using different gratings. The information listed for the Keck data is only including the 6$^{th}$ order in which the He~{\sc II}~$\lambda$5411 line is present. The S/N represented for the Keck data is for the combined spectra used in our study of the CIRs, while the 35 uncombined Keck spectra were used to observe small-scale structures (Section 4.4).}
\label{tab:tab1}
\end{center}
\end{table*}

The remaining data were collected at five professional observatories: (1) the Nordic Optical Telescope (NOT) located at Roque de los Muchachos Observatory, La Palma in the Canary Islands, (2) the Dominion Astrophysical Observatory (DAO) in British Columbia, Canada, (3) the Ond\v{r}ejov Observatory at the Astronomical Institute of The Czech Academy of Sciences in the Czech Republic, (4) the Observatoire du Mont-M\'egantic (OMM) located in Qu\'ebec, Canada and (5) the Keck Observatory located in Hawai'i, USA. All observers participating in the campaign observed at least the He~{\sc II}~$\lambda$5411 emission line where the CIR perturbations can be easily detected due to the strength of the line along with its relative isolation. Table~\ref{tab:tab1} also provides the instrument information for each observatory. Each observatory supplied the necessary bias, dark and flat fields for reduction. The wavelength calibration was accomplished by standard discharge lamps before or after each exposure. Lastly, four of us collected spectra with private instrumentation described in Table~\ref{tab:tab1}. We reduced these data with standard techniques utilizing bias, dark and flat frames obtained as close to the observations as possible. Despite the small telescope size, the data are of high quality in both signal-to-noise (S/N) and in resolution.

\begin{figure}
\begin{center}
\hspace*{-1cm}
\includegraphics[width=0.75\columnwidth,angle=90]{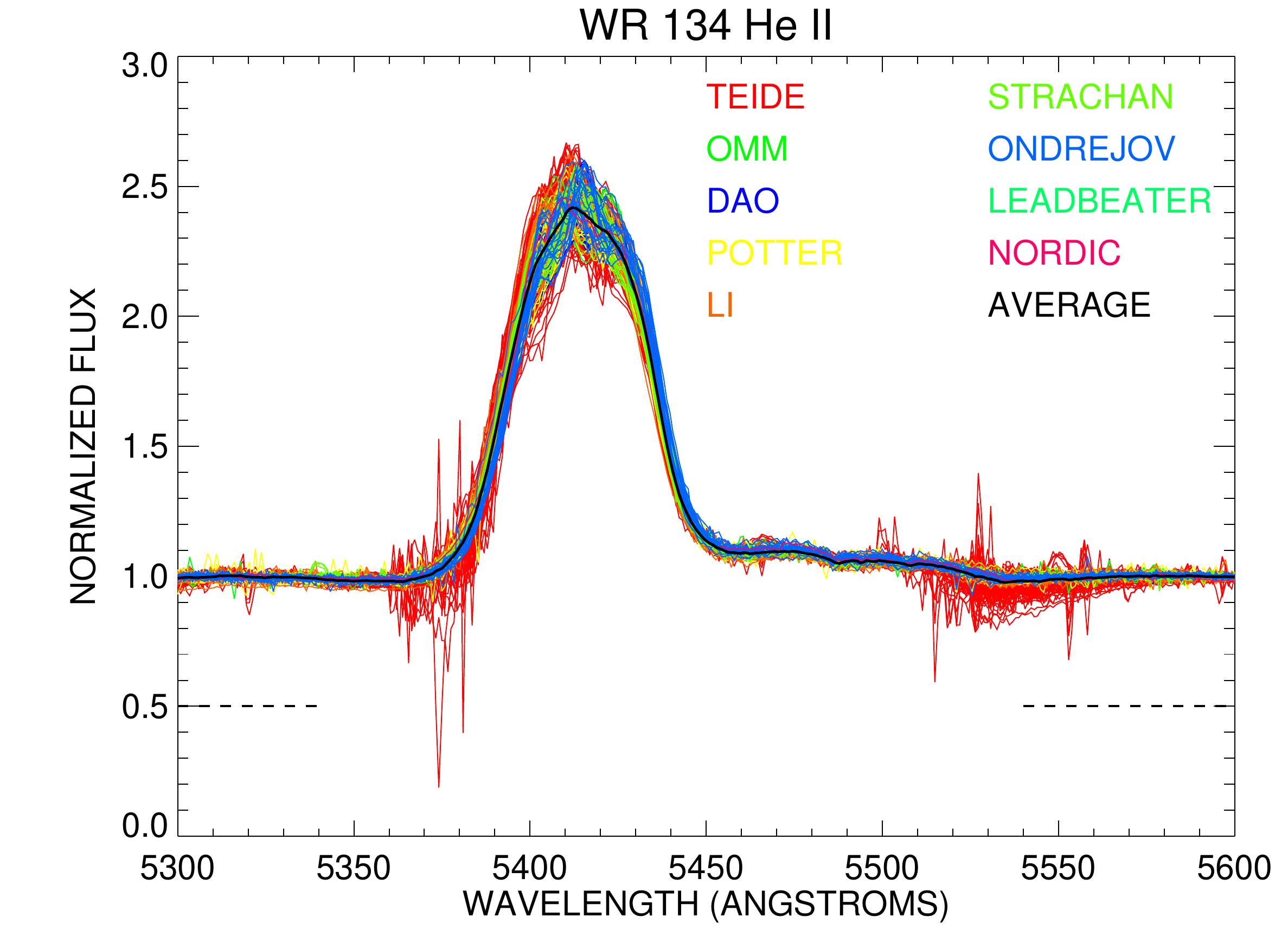}
\caption[All of the spectra obtained around He~{\sc ii} for WR 134.]{Superposition of all 395 spectra obtained around He~{\sc II}~$\lambda$5411. Different data sources/observers and an average profile are shown related to the plot legend. The noise in the Teide data (red) is due to the edges of the echelle orders. Dashed lines represent the normalized continuum regions.}
\label{fig:fig3}
\end{center}
\end{figure}

In order to compare all of the data, we corrected each spectrum for the heliocentric velocity shift from Earth and normalized each spectrum in the vicinity of the line(s) of interest. In some cases, normalization or blaze removal was aided by observations of Regulus (B8IVn) or $\zeta$ Aql (A0IV-Vnn) to correct for spectrograph response. All spectra of WR 134 were normalized around the following continuum regions: 5305 - 5340 \AA, 5540 - 5629 \AA\  and 6755 - 6775 \AA. We show the complete spectroscopic data set in Fig.~\ref{fig:fig3}. We note that the He~{\sc II}~$\lambda$5411 line is positioned on the eShel such that the line wings suffer from low S/N in several spectra. The core of the line is unaffected, allowing for a useful measurement of most quantities except for equivalent widths. Other lines were better placed and did not suffer from the noisy line wings. Further details of the data reductions are given by \citet{AldorettaThesis2015}.

\section{MEASUREMENTS}

\begin{figure}
\begin{center}
\includegraphics[width=0.95\columnwidth]{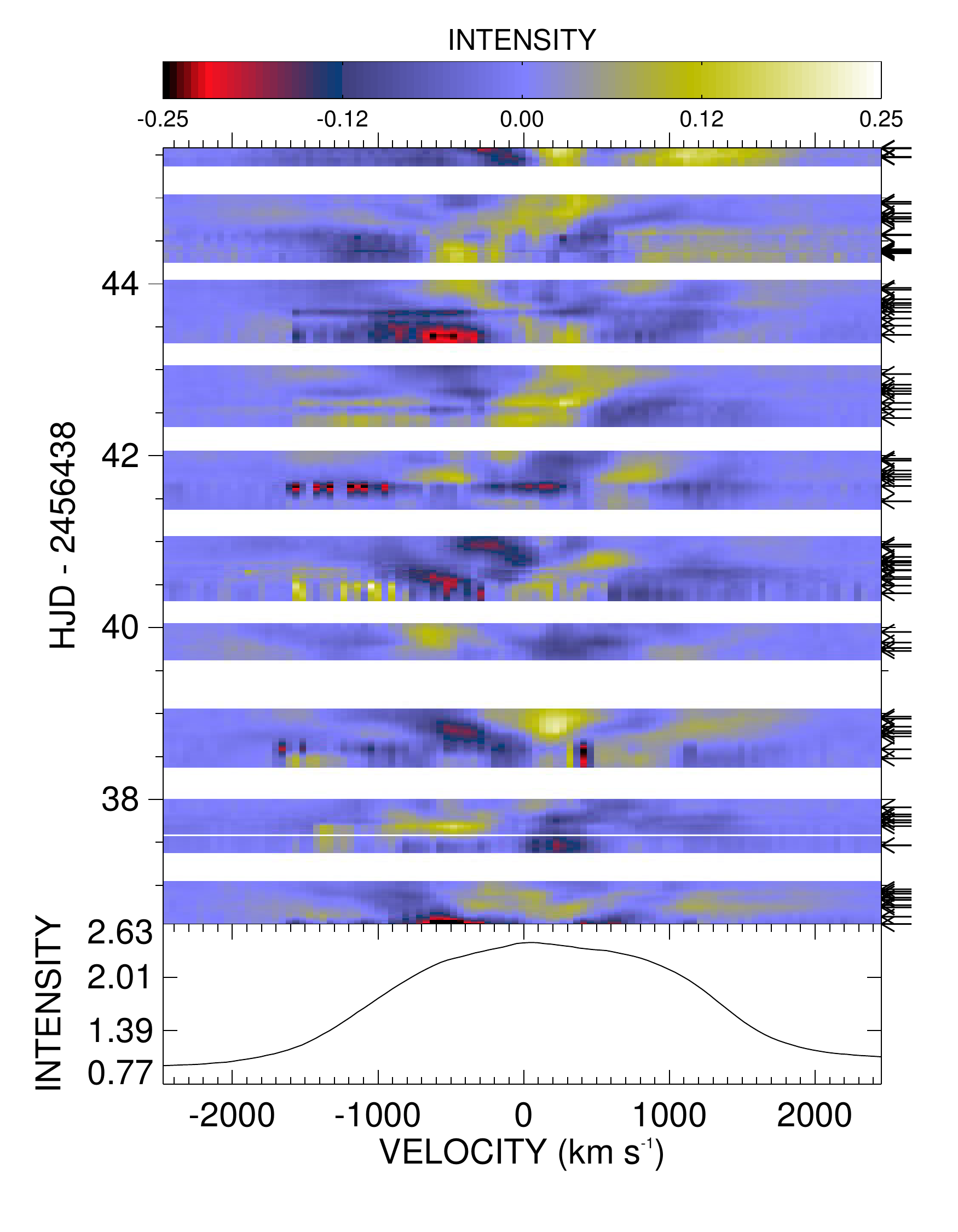}
\caption[A difference plot of the He~{\sc II} emission line from a sample of the dataset.]{A difference plot from the mean He~{\sc II}~$\lambda$5411 spectrum for a sample of the dataset. The top panel shows the intensity scale while the bottom panel shows the mean spectrum in velocity space. The arrows indicate the location of each spectrum in time. The starting HJD is the beginning of the campaign. The remaining difference plots can be found in Appendix~A, Fig.~A1 - A11 (online only).}
\label{fig:fig4}
\end{center}
\end{figure}

In their extensive study of WR 134, \citet{Morel1999} made several measurements involving the various moments of the spectral lines to characterize the spectroscopic variability. We made similar measurements on our dataset. However, before any calculations could be made, the spectra were rebinned on a common wavelength grid having a spectral resolving power of $R =$ 5,300, consistent with the lowest resolution spectra obtained during the campaign. Once all the spectra were properly rebinned, we constructed grayscale images of the differences from the mean as a function of time for the spectral region around the He~{\sc II}~$\lambda$5411 emission line. Variability is easily seen in the difference plots, as displayed in Fig.~\ref{fig:fig4}. By calculating the moments of this line, the exact period of the spectral variability can be measured on both short and long timescales. 

\subsection{Calculating Moments}

Statistical analysis plays a major role in determining the gross shapes of emission lines. Both large and small-scale variabilities that are present in WR winds have different effects on these broad lines and their symmetries over time. With evidence for the presence of a CIR having been previously found \citep{Morel1999}, the 4-month spectroscopic data were expected to show these large features moving across the top of the He~{\sc II}~$\lambda$5411 emission line. Calculating the normalized, low-order moments is an effective way to characterize the global shapes of the spectral lines. While the zeroth moment of a Gaussian gives the equivalent width (EW) measurement, two dimensionless quantities can also be calculated from these moments: the skewness and the kurtosis. The skewness defines the line asymmetry from the central wavelength and the kurtosis measures the degree to which the line is peaked. These, along with several other quantities, were calculated for each spectrum in the entire dataset for the $\lambda$5360 - 5455 $\AA$ wavelength region. To give an idea of the general level of these quantities and the range over which they vary, we calculated the means and standard deviation (i.e. scatter $\sigma$) of all of these measurements.

The mean skewness is $-$0.186 $\pm$ 0.016, while the mean kurtosis gives 2.231 $\pm$ 0.013. The equivalent width (EW), flux-weighted velocity, the bisector velocity and the full width at half maximum (FWHM) were also calculated in the same wavelength region for each spectrum. A mean value of $-$62 $\pm$ 2 \AA\ was found for the EW, and the flux-weighted velocity of the emission line gives $-$74 $\pm$ 11 km s$^{-1}$. For the bisector and FWHM calculations, the velocity at each wavelength was determined, with the rest wavelength at $v =$ 0 km s$^{-1}$. These velocity values were then linearly interpolated with the normalized flux values for both the blue side ($-$3000 to 0 km s$^{-1}$) and the red side (0 to +3000 km s$^{-1}$) of the emission line. The bisector gave a mean value over the normalized line flux interval from 1.0 to 1.4 of 143 $\pm$ 34 km s$^{-1}$. The FWHM values (mean of 2932 $\pm$ 60 km s$^{-1}$) were determined by subtracting the interpolated velocity values on the red side of the line from the values on the blue side of the line at a rectified flux level of 1.4. A complete table of these values (EW, skewness, kurtosis, flux-weighted velocity, bisector and FWHM) for each spectrum can be found in Table~\ref{tab:tab5}. A more detailed description of the calculations for each of these values is given by \citet{Chene2010}.

\begin{center}
\begin{table*}
\begin{minipage}{160mm}
\begin{tabular}{lccccccccr}
\hline 
\textbf{HJD} & 
\textbf{EW} &
\textbf{SKEW} & 
\textbf{SKEW} & 
\textbf{KURT} & 
\textbf{KURT} & 
\textbf{BISEC} & 
\textbf{FLUX VEL} & 
\textbf{FWHM} & 
\textbf{OBS} \\ 
\textbf{- 2450000} &
\textbf{(\AA)} &
\textbf{} &
\textbf{$\sigma$} &
\textbf{} &
\textbf{$\sigma$} &
\textbf{(\it km s$^{-1}$)} &
\textbf{(\it km s$^{-1}$)} &
\textbf{(\it km s$^{-1}$)} &
\textbf{} \\
\hline \hline
6438.5190 & -62.6 & -0.1759 & 0.0015 & 2.2209 & 0.0025 & 118.7 & -76.4  & 3008.8 & Teide \\
6438.5415 & -62.7 & -0.1718 & 0.0012 & 2.2184 & 0.0019 & 143.9 & -84.6  & 2913.1 & Teide \\
6440.5220 & -65.0 & -0.1586 & 0.0010 & 2.2229 & 0.0016 & 104.9 & -79.5  & 3079.6 & Teide \\
6443.6069 & -59.2 & -0.1820 & 0.0013 & 2.2280 & 0.0022 & 113.6 & -70.8  & 2928.6 & Teide \\
6444.5068 & -58.9 & -0.1914 & 0.0019 & 2.2259 & 0.0031 & 154.7 & -71.2  & 2852.3 & Teide \\
6444.5830 & -58.6 & -0.1929 & 0.0019 & 2.2245 & 0.0030 & 151.3 & -70.3  & 2848.8 & Teide \\
6445.6636 & -56.0 & -0.1758 & 0.0027 & 2.1879 & 0.0042 & 145.5 & -79.4  & 2892.0 & Teide \\
6445.6816 & -62.4 & -0.1804 & 0.0046 & 2.2222 & 0.0075 & 154.6 & -73.5  & 2972.9 & Potter \\
6445.7666 & -61.6 & -0.1910 & 0.0034 & 2.2218 & 0.0055 & 161.1 & -69.0  & 2966.3 & Potter \\
6446.5918 & -51.9 & -0.1918 & 0.0038 & 2.1842 & 0.0060 & 178.4 & -73.3  & 2758.9 & Teide \\
\hline
\end{tabular}
\caption{Sample table of values for each moment of the He~{\sc II}~$\lambda$5411 emission line of WR 134. Including the corresponding HJD (- 2450000), equivalent width, skewness and error, kurtosis and error, bisector velocity, flux-weighted velocity, full-width at half-maximum and observer. The entire table is available in Appendix~B (online only).}
\label{tab:tab5}
\end{minipage}
\end{table*}
\end{center}

\subsection{Period Search}

\begin{figure*}
\begin{center}
\hspace*{-1.2cm}
\begin{tabular}{ccc}
\includegraphics[width=0.6\columnwidth]{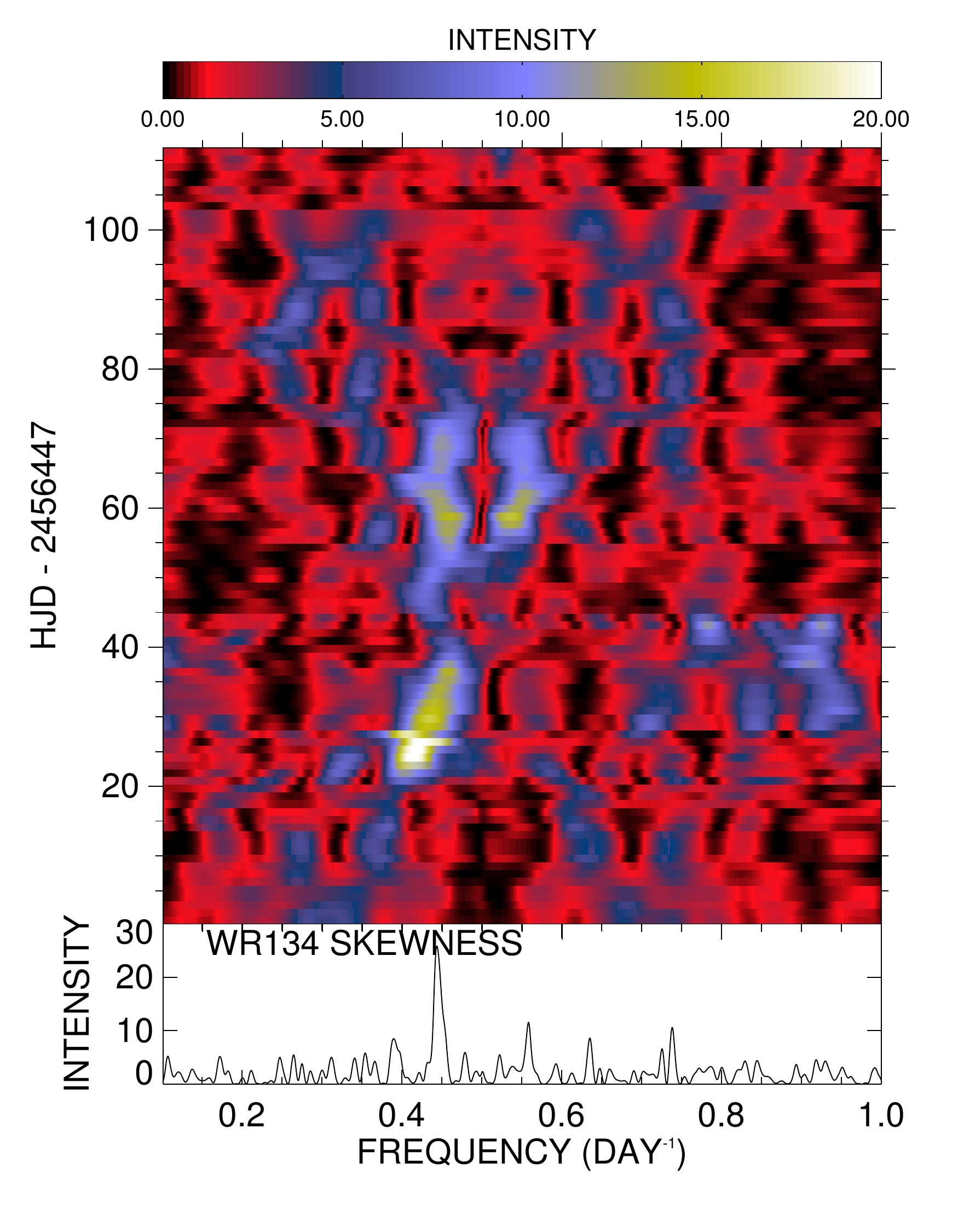} & \includegraphics[width=0.6\columnwidth]{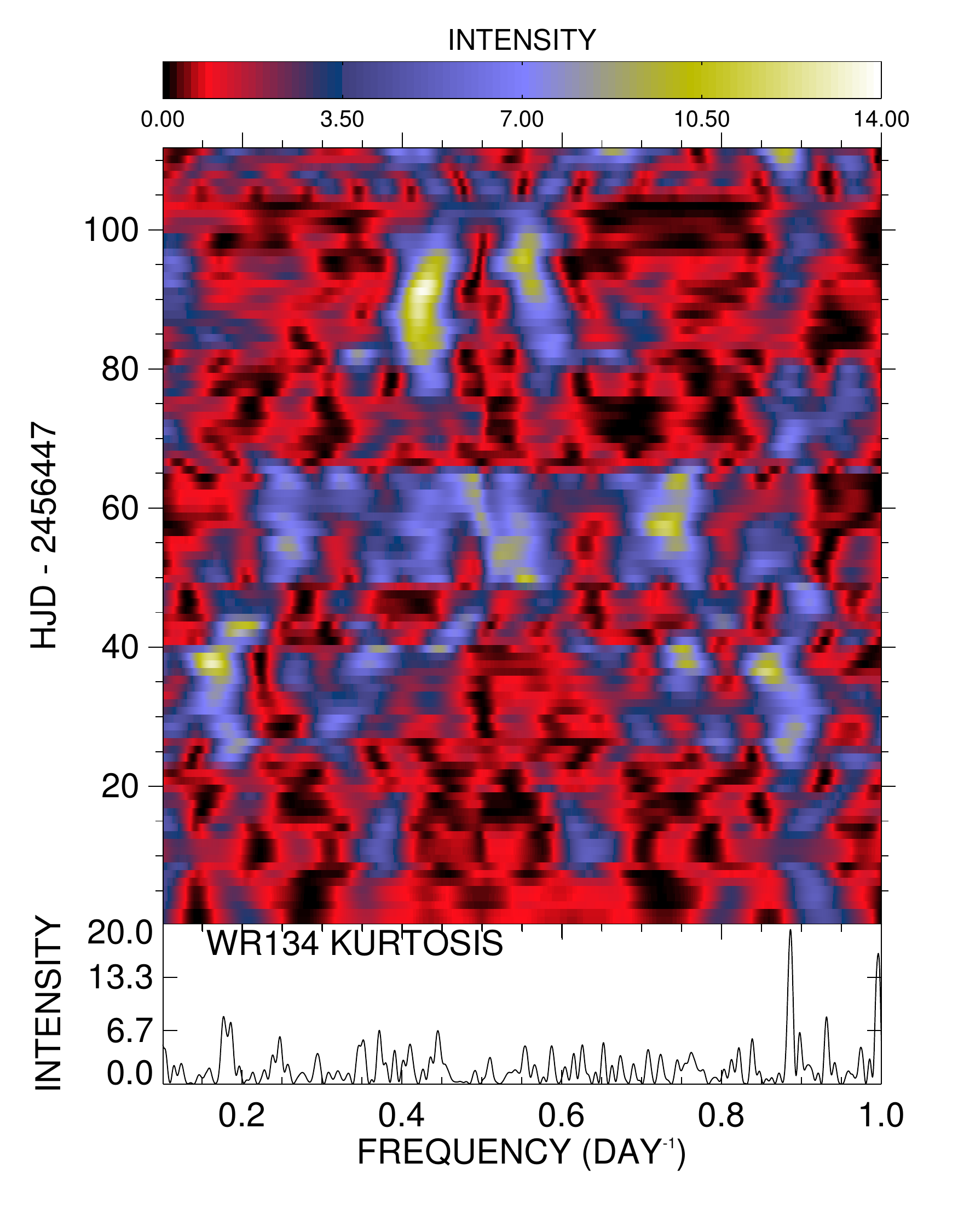} &
\includegraphics[width=0.6\columnwidth]{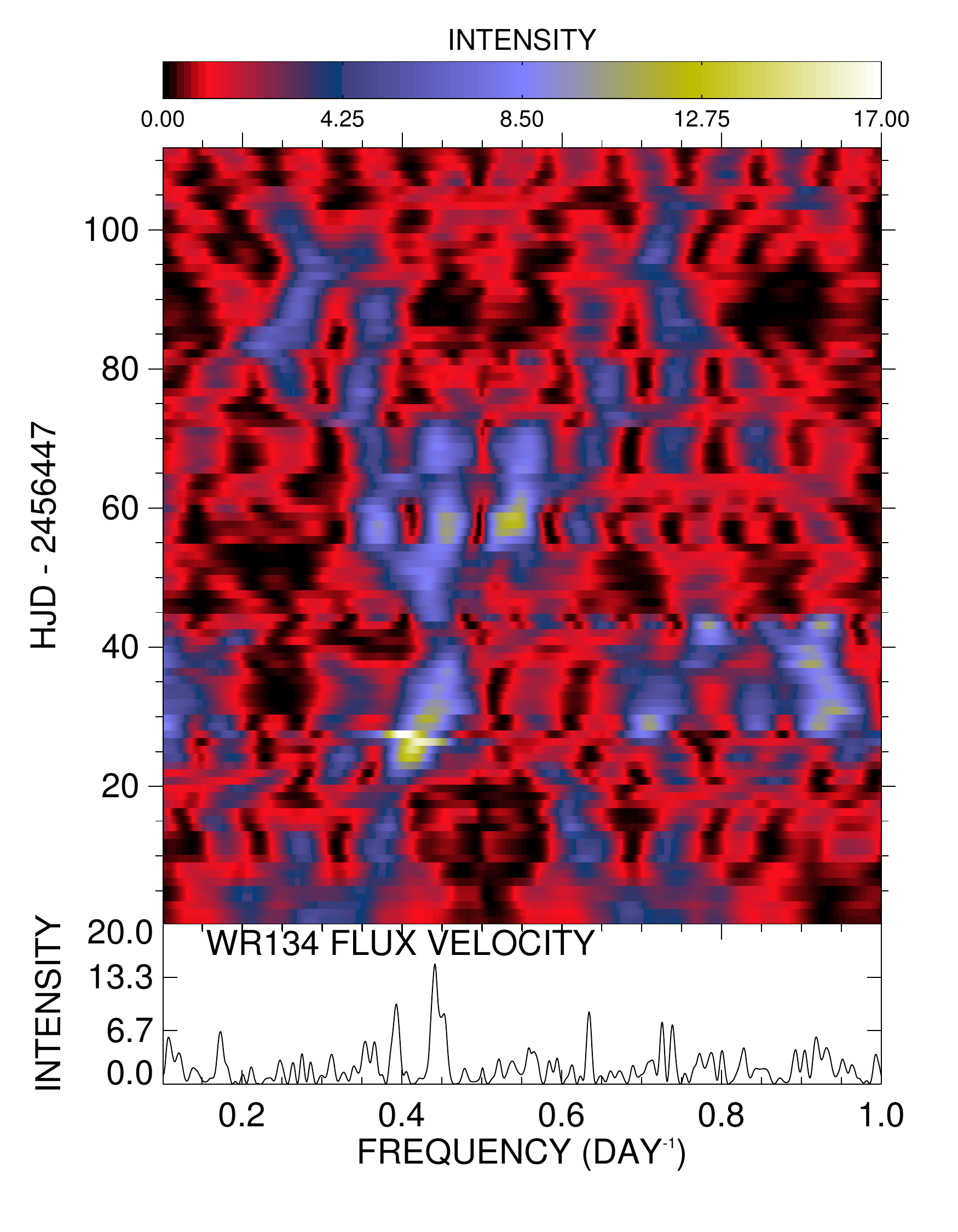} \\
\includegraphics[width=0.6\columnwidth]{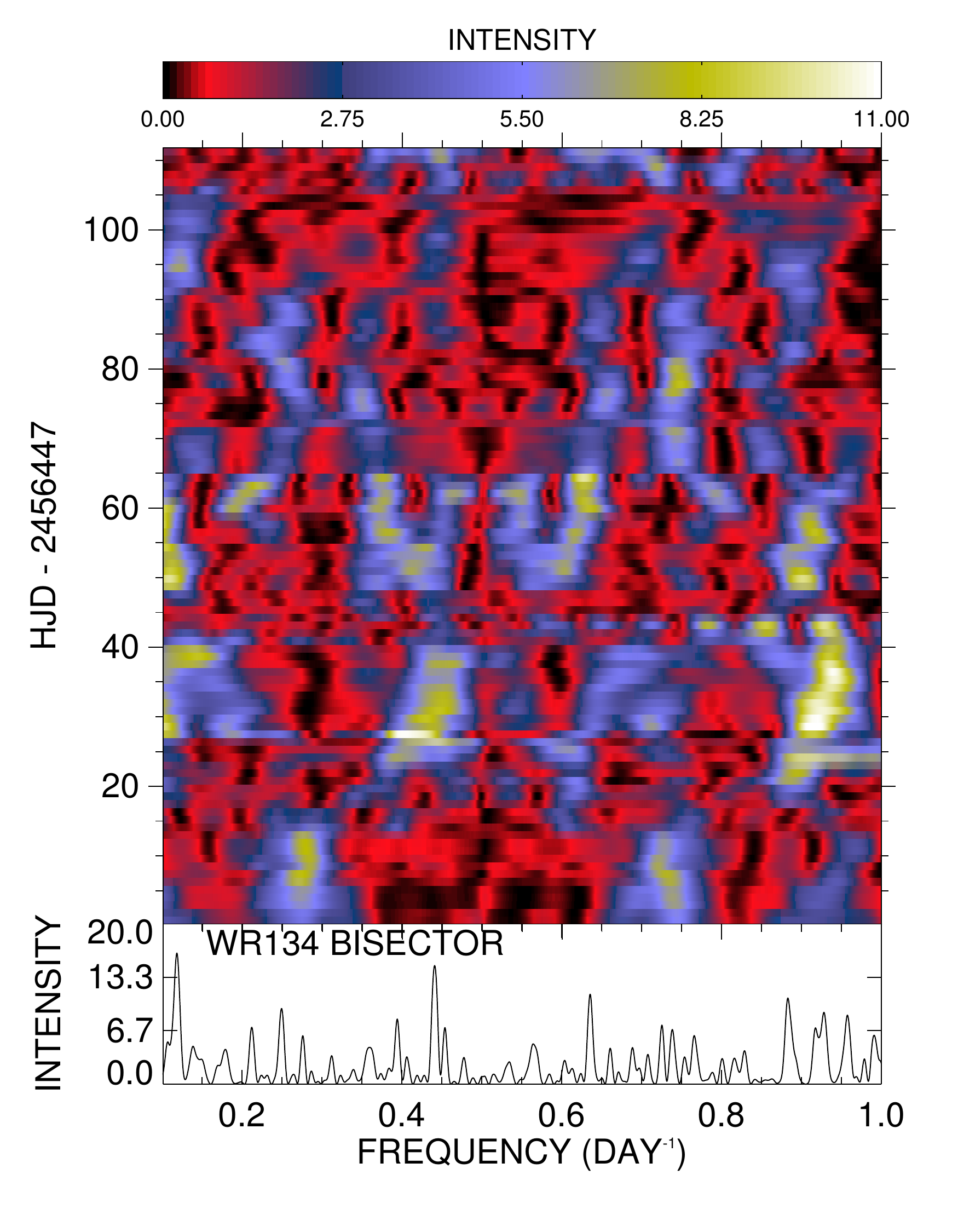} &
\includegraphics[width=0.6\columnwidth]{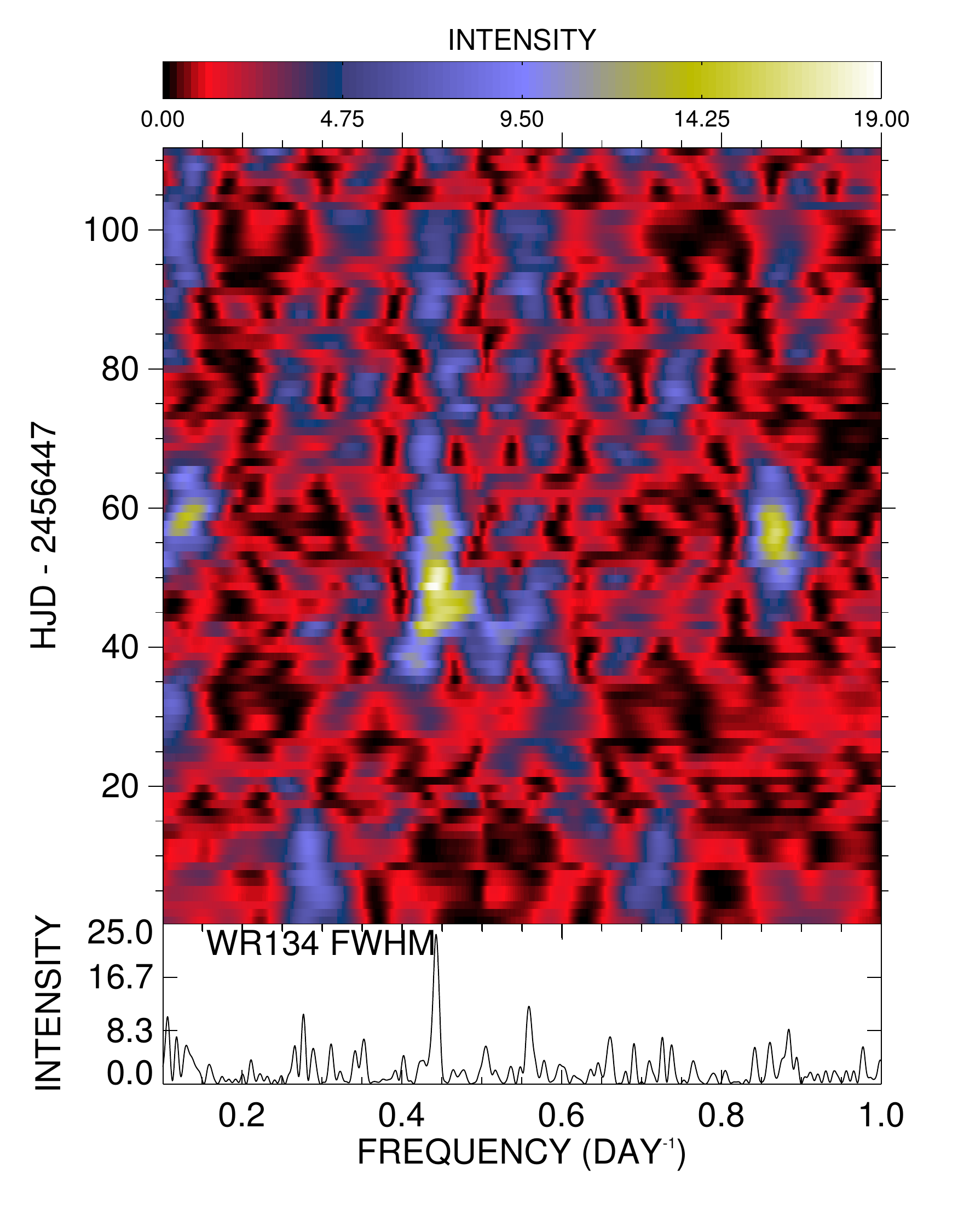} \\
\end{tabular}
\caption[Scargle analysis for five series of measurements for the He~{\sc II} emission line.]{Time-dependent Scargle analysis for the Skewness, the Kurtosis, the flux-weighted velocity, bisector and FWHM of the He~{\sc II}~$\lambda$5411 emission line. The bottom panels show the Scargle periodogram for the entire dataset while the top panels show the Scargle analysis as a function of time using 10-day bins with 20\% overlap.}
\label{fig:fig5}
\end{center}
\end{figure*}

From five of the series of measurements, a Scargle periodogram \citep{Scargle1982} was calculated to search for a periodicity within the variability of these quantities. Fig.~\ref{fig:fig5} shows these periodograms as a function of time, using 10-day intervals, for the entire dataset. While the values of the detected periods for each of the five moments are seen to fluctuate over the four month time frame, there are clear time intervals that show stronger detections. Along with a Scargle analysis, a phase determination minimization (PDM) analysis \citep{Stellingwerf1978} was also utilized. Table~\ref{tab:tab3} lists the average frequency found using these two methods for each of the five measurements calculated. The overall average period was calculated to be $P = $ 2.255 $\pm$ 0.008 (s.d.) days using all ten Scargle and PDM values. The kurtosis has a peak centred around the $2f$ harmonic of the period. We adopted the frequency similar to the other measurements for this line. The double harmonic is normal in this situation as it refers to how peaked the line is. When a perturbation moves across the line with frequency $f$ in a sinusoidal pattern, we will see it directly as $f$ with velocities and the skewness, but would cause a double sinusoidal pattern for the variability in the peak.

\begin{table}
\begin{center}
\begin{tabular}{|c c c|}
\hline
\textbf{Moment} & \textbf{Scargle analysis} & \textbf{PDM analysis} \\ [1ex]
  & \textbf{(day$^{-1}$)} & \textbf{(day$^{-1}$)} \\ [1ex]
\hline
Skewness & 0.4452 $\pm$ 0.008 & 0.4451 $\pm$ 0.005 \\
Kurtosis & 0.4432 $\pm$ 0.007 & 0.4450 $\pm$ 0.006 \\
Flux Velocity & 0.4449 $\pm$ 0.011 & 0.4437 $\pm$ 0.007 \\
Bisector & 0.4418 $\pm$ 0.009 & 0.4409 $\pm$ 0.010 \\
FWHM & 0.4427 $\pm$ 0.006 & 0.4428 $\pm$ 0.007 \\ [0.5ex]
\hline
\end{tabular}
\caption[The list of the average measurements for Scargle and PDM analysis.]{List of the five measurements for both a \citet{Scargle1982} and PDM \citep{Stellingwerf1978} analysis. All errors are given in standard deviation.}
\label{tab:tab3}
\end{center}
\end{table}

\begin{figure}
\begin{center}
\includegraphics[width=0.7\columnwidth,angle=90]{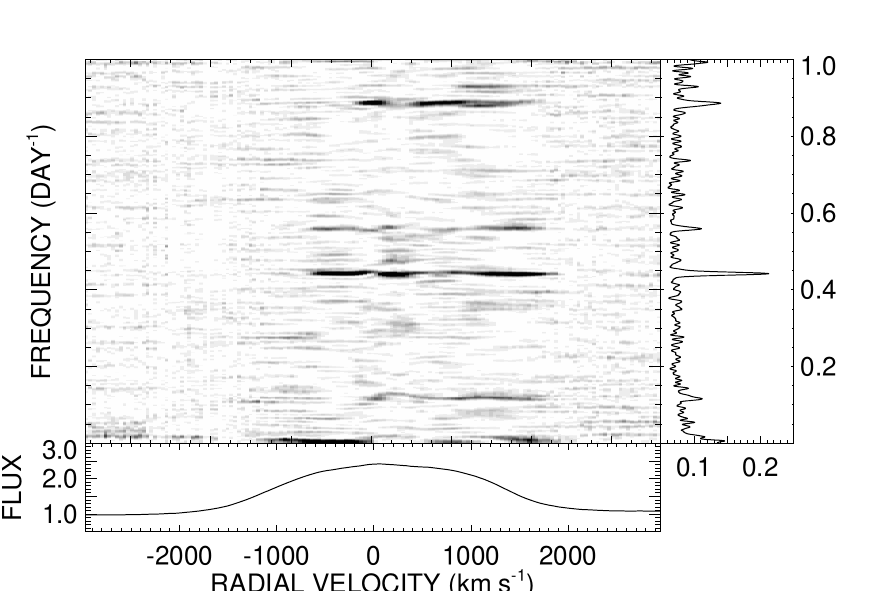}
\caption[The 2-D Scargle peridogram around the He~{\sc II} emission line.]{The 2-D \citet{Scargle1982} periodogram around He~{\sc II}~$\lambda$5411 line. The bottom panel shows the average spectrum of the entire dataset in velocity space while the right panel shows the Scargle periodogram averaged over the entire line. The main panel shows the grayscale of the Scargle analysis for each velocity bin. It can be seen that the frequency of $\sim$ 0.4435 is the strongest peak.}
\label{fig:fig6}
\end{center}
\end{figure}

\begin{figure}
\begin{center}
\includegraphics[width=0.95\columnwidth]{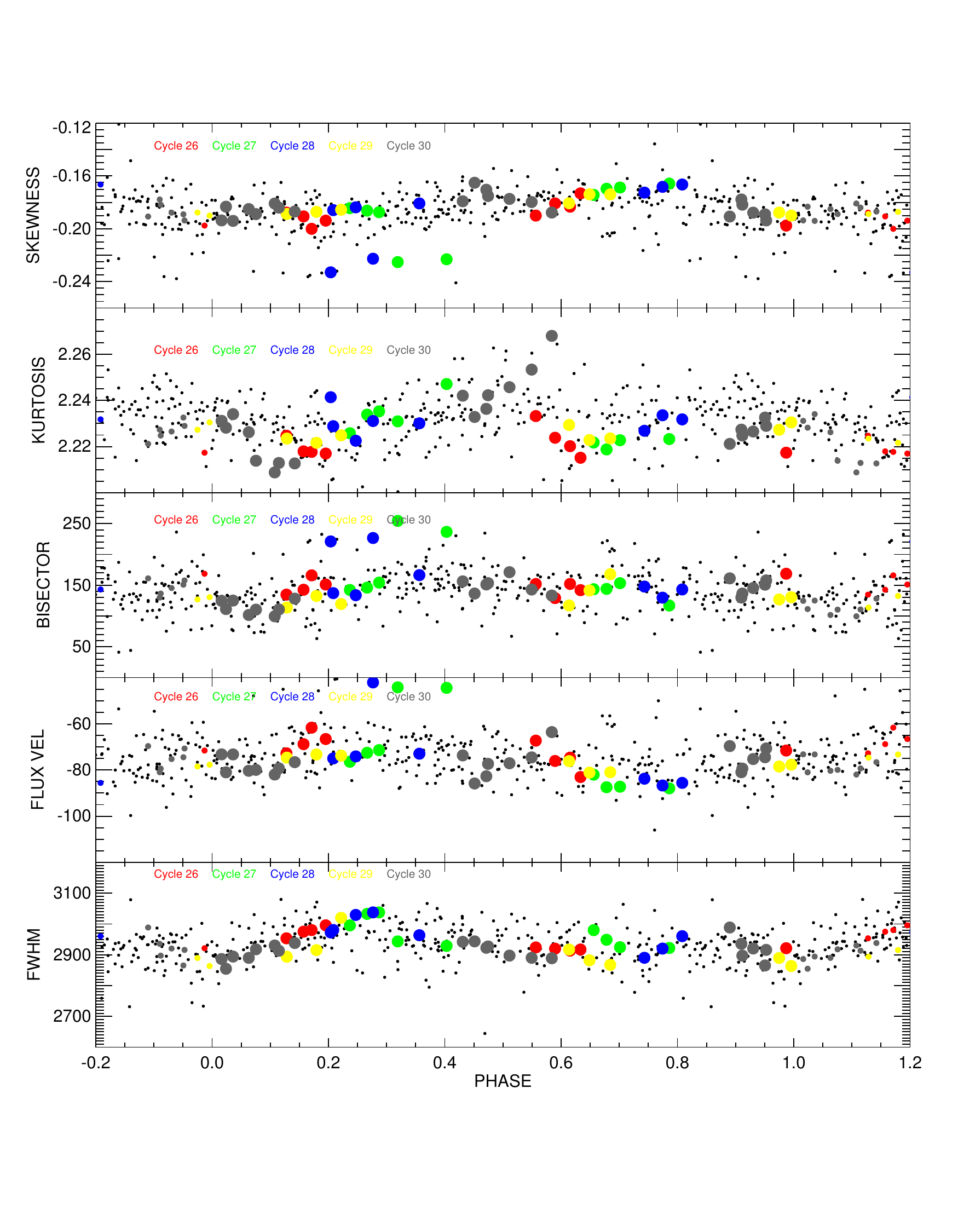}
\caption[The phased measurements using the 2.255-day period.]{The phased measurements for selected cycles using the 2.255-day period. Each panel shows all 395 measurements as small black dots for each moment, while also showing certain cycles as coloured dots (see legends). The smaller coloured dots represent the data at overlapping phases. The remaining phased moments for the entire dataset can be found in Appendix~A, Fig.~A12 - A22 (online only).}
\label{fig:fig7}
\end{center}
\end{figure}

\begin{figure}
\begin{center}
\hspace*{-1cm}
\includegraphics[width=0.65\columnwidth,angle=90]{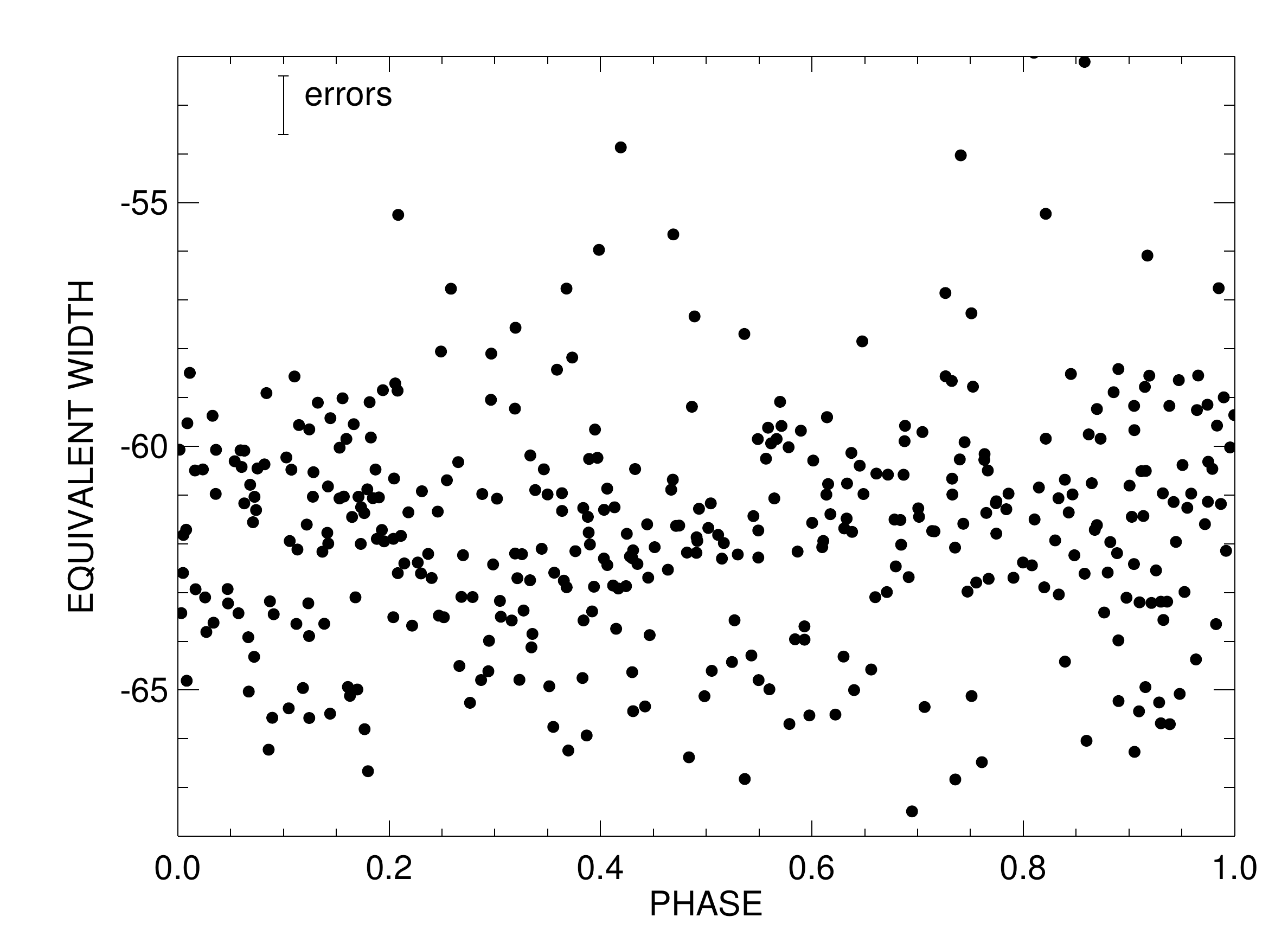}
\caption[The equivalent width measurements of the He~{\sc II} emission line as a function of phase.]{The EW measurements of the He~{\sc II}~$\lambda$5411 emission line plotted as a function of phase using the 2.255-day period for all spectra in our dataset. The average calculated error for these values is displayed.}
\label{fig:fig8}
\end{center}
\end{figure}

Once the overall period was determined using the measurements of the He~{\sc II}~$\lambda$5411 emission line, we calculated a 2-dimensional Scargle periodogram \citep{Scargle1982} for this line in velocity space for the entire data set. This analysis shows that a frequency of $\sim$ 0.4435 day$^{-1}$ ($P =$ 2.255 days) has the strongest peak across most of the profile, as can be seen in Fig.~\ref{fig:fig6}. Along with the strongest fundamental frequency, both its first harmonic ($2f$) and their aliases, ($1 - f$) and ($1 - 2f$) due to a $1$-day sampling, have strong peaks as well.
\citet{Morel1999} created a similar CLEANed 2-dimensional power spectrum (PS) around the He~{\sc II}~$\lambda$4686 emission line. Their results showed that their highest frequency peak was $\sim$ 0.444 day$^{-1}$ ($P = $ 2.25 $\pm$ 0.05 days). We were able to confirm this period within the errors to a more precise value. Each of the five sets of measurements were then phased according to our 2.255-day period, with $t_0$ being the first day of the campaign (HJD 2456438) as the arbitrarily assigned origin of the phases as no obvious origin point, such as periastron passage, exists in this analysis. To illustrate the phase-dependent behaviour of the five measured quantities, we plot in Fig.~\ref{fig:fig7} measurements from five consecutive cycles in our dataset. It can be seen that the skewness values display a single sine curve with phase, as expected for one sub peak moving from one side of the profile to the other. Fig.~\ref{fig:fig7} also shows the expected double sine curve for kurtosis, as mentioned previously. The equivalent width (EW) values were also calculated for this emission line and the scatter is clearly above the typical error bar \citep{Vollmann2006}. The measurements did not show any trend with phase, as can be seen in Fig.~\ref{fig:fig8}. 

\section{RESULTS}

\begin{figure}
\begin{center}
\hspace*{-1cm}
\includegraphics[width=0.95\columnwidth]{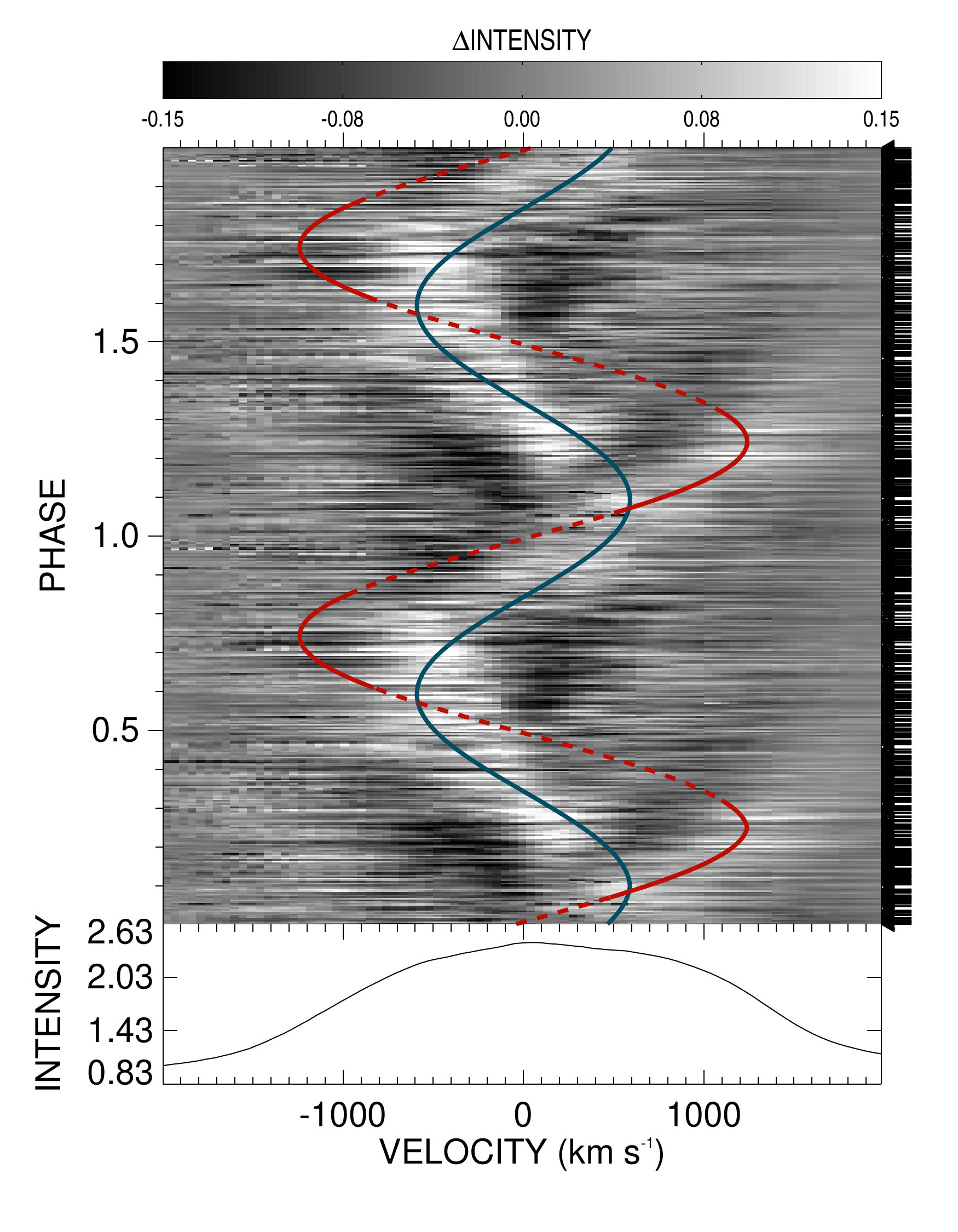}
\caption{The global difference grayscale image of the He~{\sc II}~$\lambda$5411 emission line for the entire campaign. The blue line represents the kinematic model calculated in Section 4.2 of the central CIR trace, while the solid and dashed red line represents the CIR trace detected with a phase shift of $\Delta \phi \sim$ 0.25. The solid red line shows where we can actually detect the CIR trace.}
\label{fig:fig9}
\end{center}
\end{figure}

Along with the moments, the spectra for the entire dataset were phased on the 2.255-day timescale. There is a strong CIR signature or trace present within the He~{\sc II}~$\lambda$5411 emission line, spanning a velocity range of $-$600 to 600 km s$^{-1}$ and describing a single sinusoid per cycle (hereafter referred to as the "central" CIR). However, there also appears to be another CIR trace present within the He~{\sc II}~$\lambda$5411 emission line, which extends out to $\sim$ 1500 km s$^{-1}$, mainly present between phases 0.00 - 0.35, which can be seen in Fig.~\ref{fig:fig9}. We have made the line solid only where we can actually see the bright (or dark) trace in the grayscale plot. This feature and the central CIR reach peak velocities at different phases with a shift of $\Delta \phi \sim$ 0.25, which we interpret as indicating that they are, in fact, located at different longitudes on the star with a separation of $\Delta \phi \simeq$ 90$^{\circ}$. Note that because we are using differences from the mean in these grayscale plots, the sum over phase or time of the difference flux at each wavelength will always equal zero. This leads to a shadowing effect, where each bright trace is accompanied by an associated dark trace, which gives the plot a distinct appearance. In the next section, we present an analysis conducted on the dataset to search for changes within the phased grayscale image with time.

\subsection{Cross-Correlation Analysis}

\begin{figure}
\begin{center}
\includegraphics[width=0.75\columnwidth,angle=90]{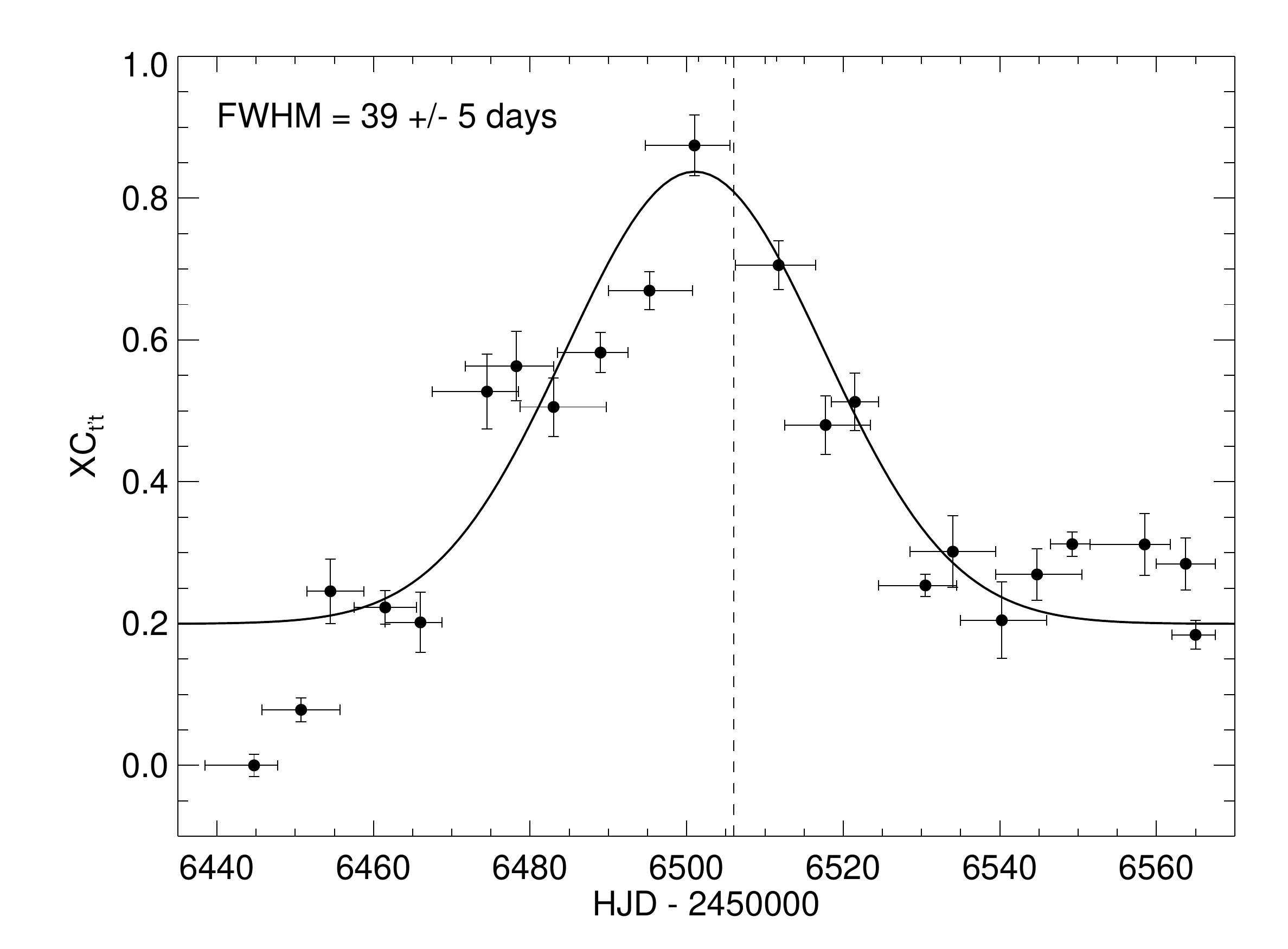} 
\caption[The cross-correlation analysis for the difference images of the He~{\sc ii} emission line.]{The cross-correlation analysis for the center reference image ({\it dashed vertical line}). The x-axis error bars show the time covered by each image, while the y-axis error bars display the $XC_{t't}$ errors. The solid line is the Gaussian function that was fitted for the analysis. The remaining two cross-correlation images are located in Appendix~A, Fig.~A23 - A25 (online only).}
\label{fig:fig10}
\end{center}
\end{figure}

In order to determine if any changes occurred within the trace of the CIRs over the timespan of the campaign, a cross-correlation analysis was conducted on several phased grayscale images. For every 5 days ($\sim$ 2 cycles), a new phased grayscale image was created, for a total of 23 for the entire data set. Three arbitrary grayscale reference images were chosen based on good phase coverage, from which a normalized 2-dimensional cross-correlation function (CCF) was calculated from the images before and after the reference image. The correlations were quantified using
\begin{equation}
XC_{t't} = {{\Sigma_{jk} I_{t'jk} \times I_{tjk}}\over{\Sigma_{jk} I_{t'jk}^2}},
\label{eq:eq6}
\end{equation}
where $I_{tjk}$ is the intensity read in the $jk$ pixel at time $t$, $t'$ is the reference image, and $j$ and $k$ refer to the pixel locations. Fig.~\ref{fig:fig10} displays the value of $XC_{t't}$ as a function of time using one of the representative difference reference images.

Similar results were seen for reference images at the beginning or end of the campaign, although these reference images suffer from less internal phase coverage. Once the cross-correlation analysis was carried out for each reference image, a Gaussian function plus a constant background was fit to the ensemble of data for a given reference image, ignoring the artificial correlation of the $t'$ image with itself (i.e. for $t = t'$). The average FWHM for the three separate Gaussian fits is 40 $\pm$ 6 days ($\sim$ 18 cycles). The cross-correlation analysis that was carried out using the mid-campaign reference image was given more weight in this calculation due to its greater intensity. This value can be interpreted as the length of time for which the variability pattern, as seen in the grayscale plots of the difference spectra, remains coherent. After this time, these grayscale plots look significantly different and are considered uncorrelated.

This is analogous to a one-dimensional cross-correlation often used in spectroscopy for measuring radial velocities for binary stars or planets. In that case, either a good target spectrum or a model spectrum is chosen as a reference for the cross-correlation, resulting in either absolute measurements if the reference was a model spectrum, or relative measurements if the reference was an individual observation. The resulting orbit will only differ in reference to the line-of-sight velocity in the final calculation. In this situation, we see that the FWHM of the Gaussian fit of the resulting measurements is always on the order of $\sim$ 40 days, which we interpret as a coherence time for a CIR pattern on a grayscale image and thus of the CIR in the wind itself. The example cross-correlations for the beginning and end of the campaign references are shown in Appendix A, Fig.~A23 - A25 (online only).

\begin{figure*}
\begin{center}
\includegraphics[width=0.69\columnwidth]{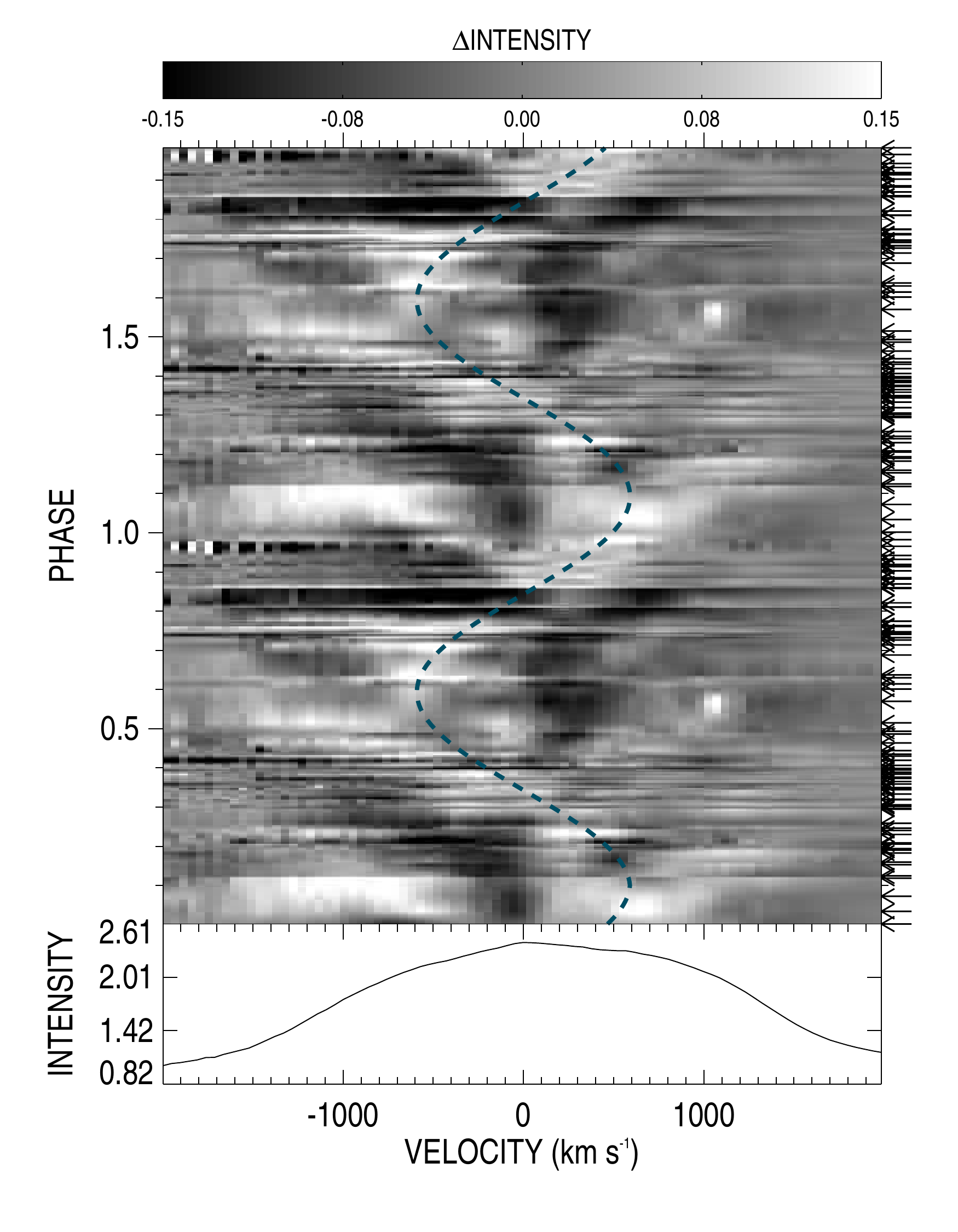} 
\includegraphics[width=0.69\columnwidth]{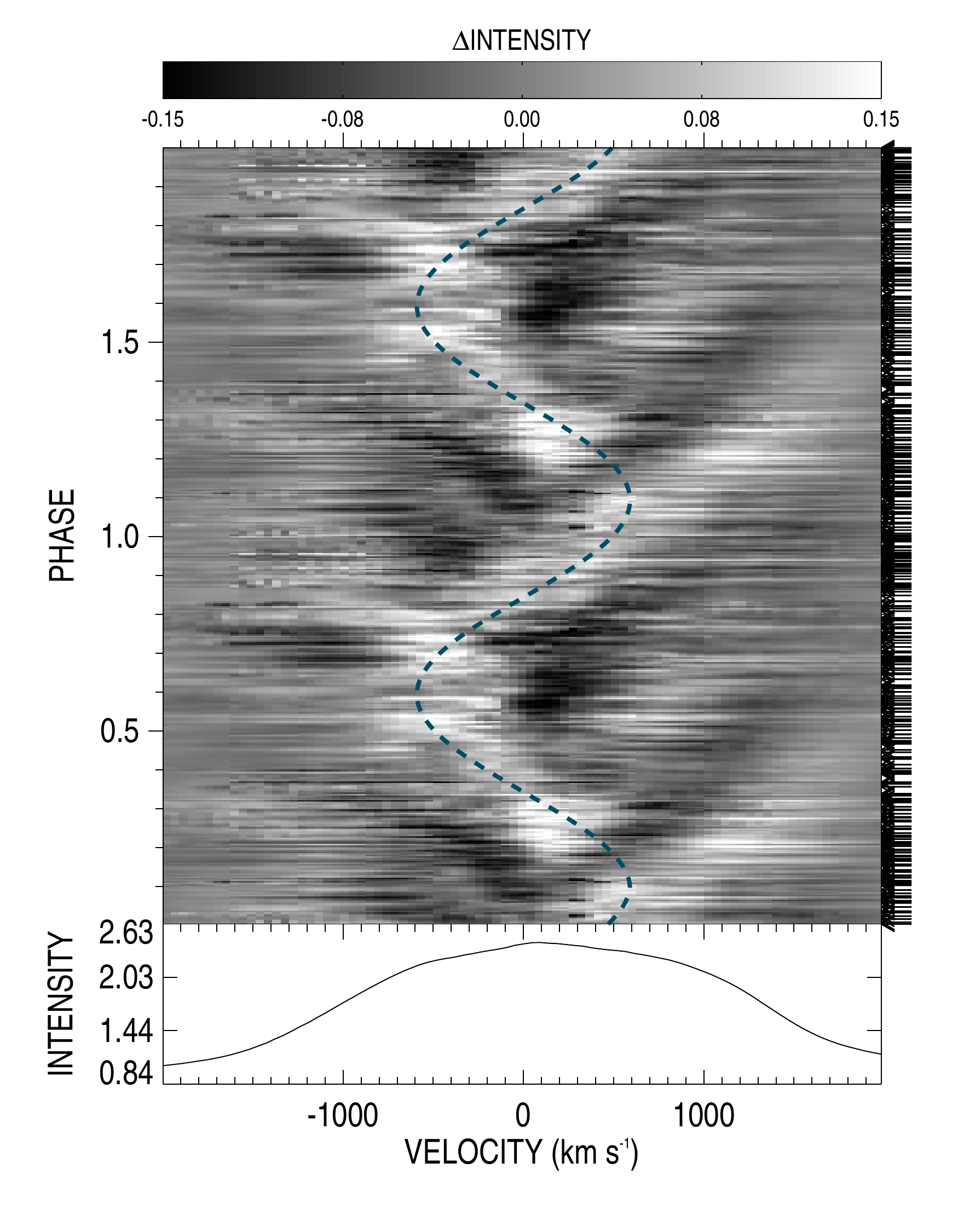} 
\includegraphics[width=0.69\columnwidth]{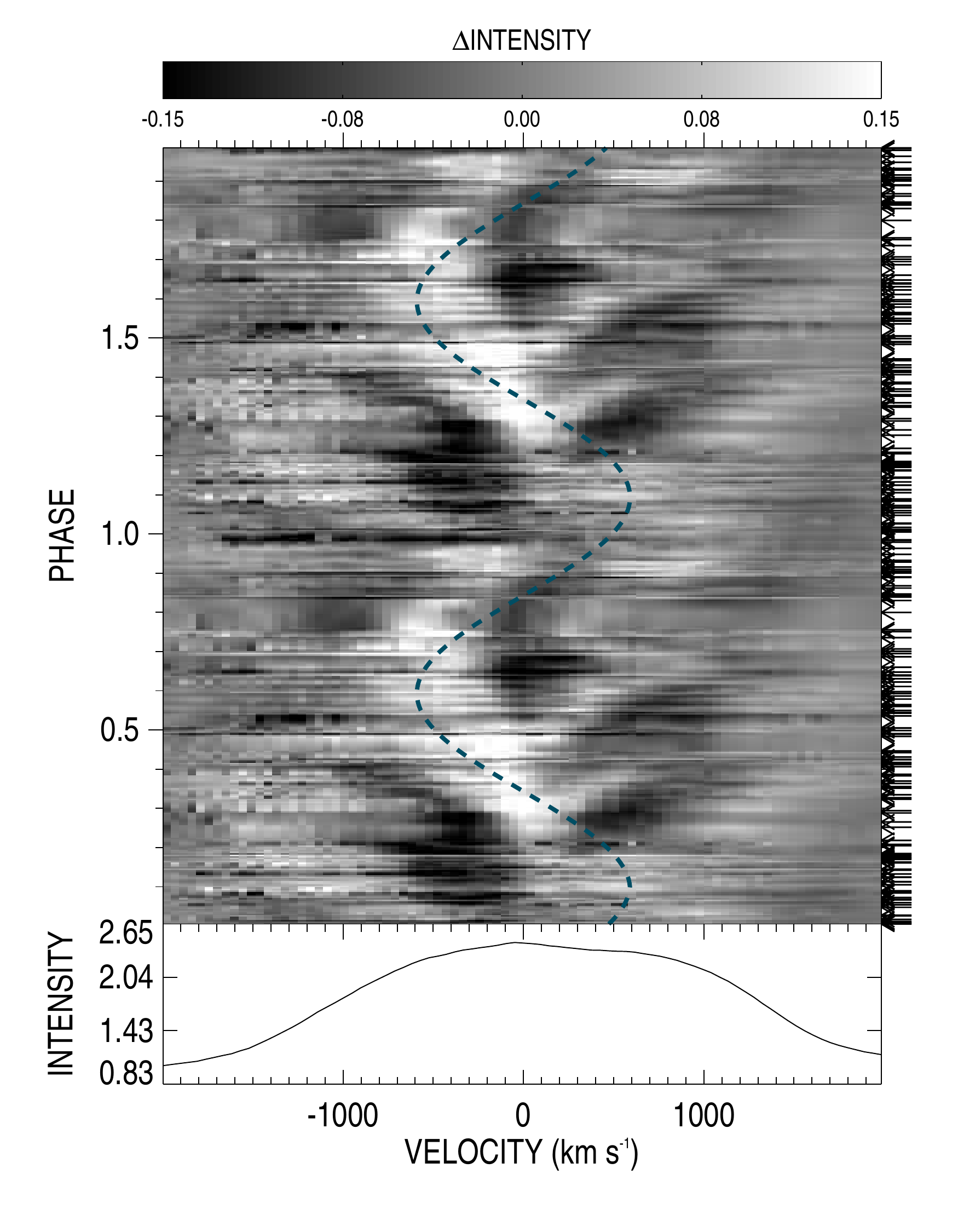}
\caption[The difference grayscale images for the first ({\it left}), second ({\it center}) and the third ({\it right}) part of the entire observing run of the He~{\sc II} emission line with the kinematics model.]{The difference grayscale images for the first ({\it left}, HJD 2456438 - 2456478), second ({\it center}, HJD 2456479 - 2456519) and the third ({\it right}, HJD 2456520 - 2456568) part of the entire observing run of the He~{\sc II}~$\lambda$5411 emission line are shown. The dashed blue line represents the kinematics model calculated in section 4.2.}
\label{fig:fig14}
\end{center}
\end{figure*}

Following this result, a set of phased grayscale difference images was created for the dataset in 40-day segments for a total of 3 images. These images are presented in Fig.~\ref{fig:fig14}, which illustrates clearly the coherence timescale of $\sim$ 40 days, as the three images have different aspects. For example, one can easily see the trace of the main CIR in the central image, but for the last 40-days of the campaign, it has a much weaker presence in the He~{\sc II}~$\lambda$5411 line. 

\subsection{Kinematics of the CIR}

In order to better understand the geometry of the central CIR in the wind of WR 134, we measured the velocity of the excess emission feature in the difference spectra used to generate the central grayscale image of Fig.~\ref{fig:fig14} (HJD 2456478 - 2456518, central third of entire run). First, a total of 20 new difference spectra were created by binning the original difference spectrum to increase the S/N of each spectrum (see Table~\ref{tab:tab4} for number of spectra in each bin). Then a Gaussian function was fit to each binned spectrum with the rest velocity of this transition used as the zero point for each fit. Each of these Gaussian fits established the velocity of the peak of the difference spectrum, indicating where the CIR excess emission was present within the emission line. In order to distinguish between the multiple CIR signatures when present, each peak was chosen by eye using the second grayscale image in Fig.~\ref{fig:fig14} to ensure the correct velocities were calculated. Table~\ref{tab:tab4} lists each velocity that was determined using this method, along with the corresponding phase, the number of spectra per bin and the weight of each bin (based on the inverse variances of the velocity errors determined by the Gaussian fitting). 
\begin{table}
\begin{center}
\begin{tabular}{|c c c c|}
\hline
\textbf{Phase} & \textbf{V$_{obs}$} & \textbf{N$_{spec}$/bin} & \textbf{Weight}\\ [1ex]
\hline
0.025 & +952.8 & 8 & 0.05 \\
0.072 & +415.6 & 8 & 0.18 \\
0.124 & +581.5 & 12 & 0.11 \\
0.172 & +264.8 & 11 & 0.15 \\
0.225 & +38.0 & 9 & 0.04 \\
0.278 & -604.1 & 7 & 0.07 \\
0.324 & +64.7 & 8 & 0.05 \\
0.380 & -437.8 & 9 & 0.12 \\
0.423 & -580.2 & 8 & 0.06 \\
0.474 & -493.5 & 9 & 0.15 \\
0.529 & -484.7 & 10 & 0.38 \\
0.624 & -509.1 & 7 & 0.09 \\
0.678 & -539.9 & 9 & 0.12 \\ 
0.725 & -47.4 & 6 & 0.07 \\
0.770 & +315.2 & 10 & 0.05 \\
0.826 & +300.4 & 7 & 0.12 \\
0.876 & +156.5 & 8 & 0.04 \\
0.927 & +693.1 & 12 & 0.06 \\
0.972 & +732.9 & 11 & 0.08 \\ [0.5ex]
\hline
\end{tabular}
\caption{The list of the phases, observed velocities (km s$^{-1}$), number of spectra in each bin and the weight of each bin for calculating the kinematics model.}
\label{tab:tab4}
\end{center}
\end{table}

\begin{figure}
\begin{center}
\hspace*{-1cm}
\includegraphics[width=0.75\columnwidth,angle=90]{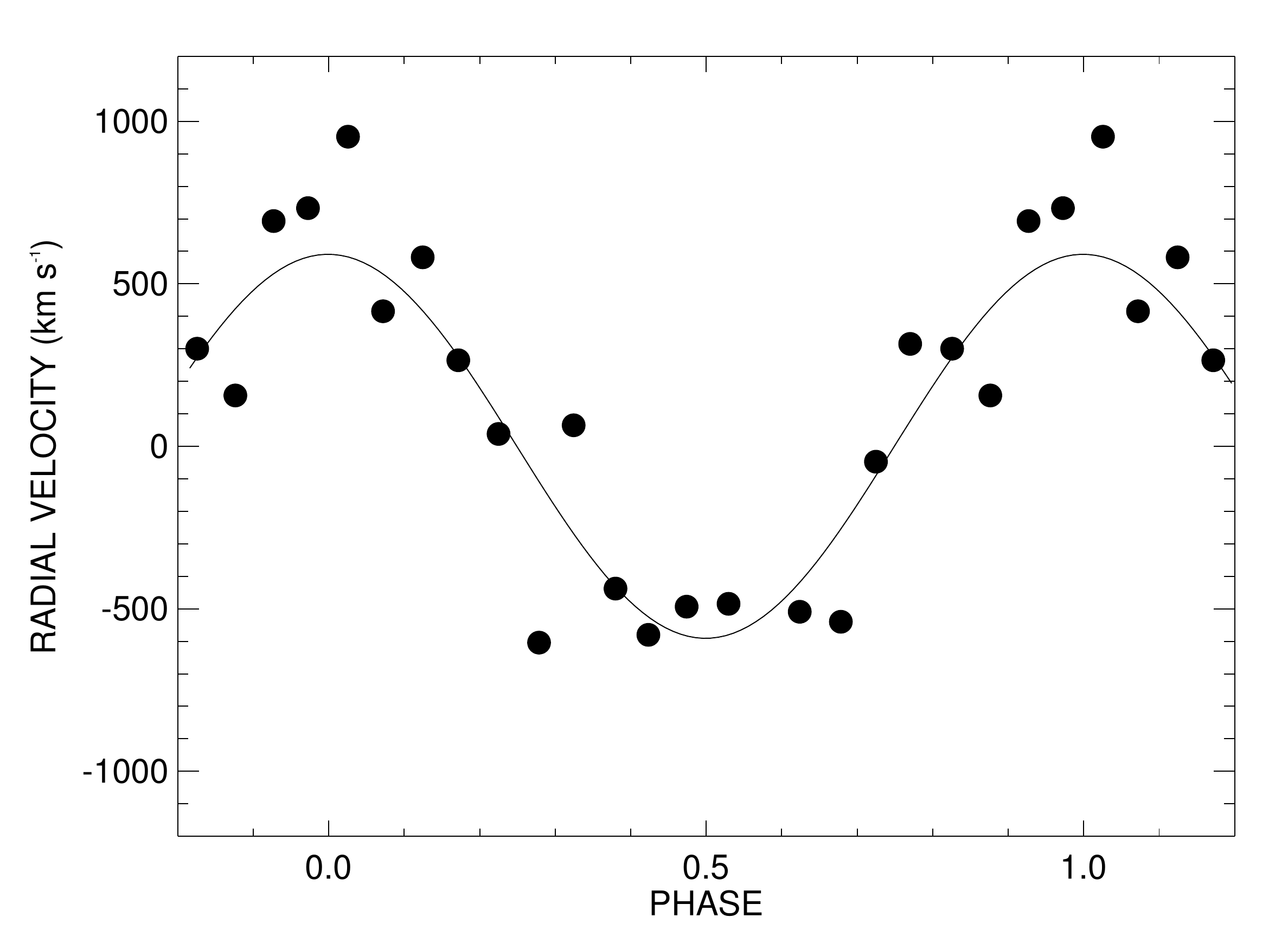}
\caption[The radial velocity of the central CIR as a function of phase.]{Radial velocity versus phase for the central CIR trace seen during one 40-day segment in Fig.~\ref{fig:fig14}. The points represent the observed velocities calculated for the binned difference spectra with the sinusoidal fit overplotted.}
\label{fig:fig13}
\end{center}
\end{figure}

Once the velocities of the excess emission from the central CIR were determined for these binned spectra, a sinusoidal curve was fit to the data, showing how the apparent velocity of the extra peak changes with phase as the star rotates. Fig.~\ref{fig:fig13} shows this fitted curve with the binned spectra, while Fig.~\ref{fig:fig14} shows the curve overplotted on the difference grayscale images. If $i$ is the angle between the stellar rotation axis and the line of sight and $\beta$ is the angle between the rotation axis and the CIR, assuming it emerges radially, then the angle between the velocity vector of the gas in the CIR and the observer at time $t$ is
\begin{equation}
\cos \theta(t) = \cos i \cos \beta + \sin i \sin \beta \cos\left({{2 \pi t}\over{P}}\right) ,
\label{eq:eq9}
\end{equation}
where $P$ is the rotation period and where we have assumed that the maximum velocity is reached at integer multiples of $P$. 

At any given time, the velocity of the excess emission peak will be $v_{e(t)} = v_{CIR} \cos \theta(t)$. Note that $v_{CIR}$ is not necessarily the terminal velocity of the wind. Indeed, as predicted by \citet{Cranmer1996}, for a bright spot the velocity of the gas in a CIR, although mainly radial, can be considerably smaller than the unperturbed wind velocity. Inspection of Fig.~\ref{fig:fig9} leads us to several conclusions. First, since the maximum velocity of the two CIR traces we detected are different, they either emerge at different angles from the rotation axis or they have a different velocity ($v_{CIR}$). Also, the median velocity of the main CIR pattern is very close to $v = 0$. This indicates that either this CIR is located at $\beta$ = 90$^{\circ}$ or that the inclination of the rotation axis is 90$^{\circ}$. Note finally that as mentioned previously, because the maximum velocity of these two CIR traces is not reached at the same phase, we can conclude that they do not emerge from the same longitude on the star. From the phase difference, we found that they are separated by $\simeq$ 90$^{\circ}$. 

\citet{Dessart2002} used the radiation hydrodynamical model of \citet{Cranmer1996} to generate synthetic emission-line profiles resulting from two CIRs created as a consequence of the presence on the surface of the star of two bright spots located at opposite longitudes at the base of an optically thin, radiatively-driven stellar wind. Although the \citet{Dessart2002} model was reproduced for a system with two CIRs with a 180$^{\circ}$ separation, our difference image clearly shows a similar brightness pattern, strongly suggesting that this is a reasonably good model to represent the observations. 

\subsection{Variations in Other Spectral Lines}

\begin{figure*}
\begin{center}
\includegraphics[width=0.69\columnwidth]{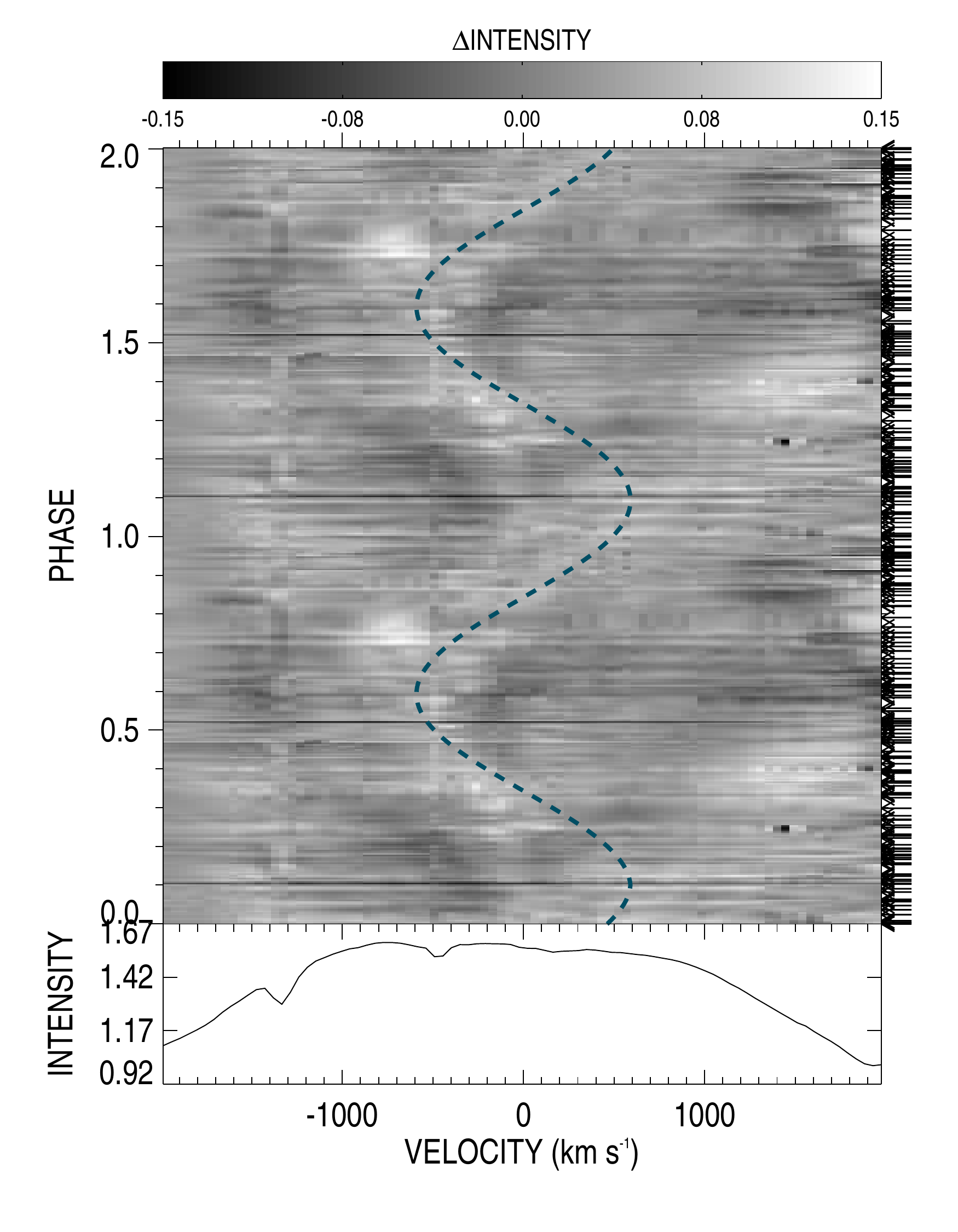}
\includegraphics[width=0.69\columnwidth]{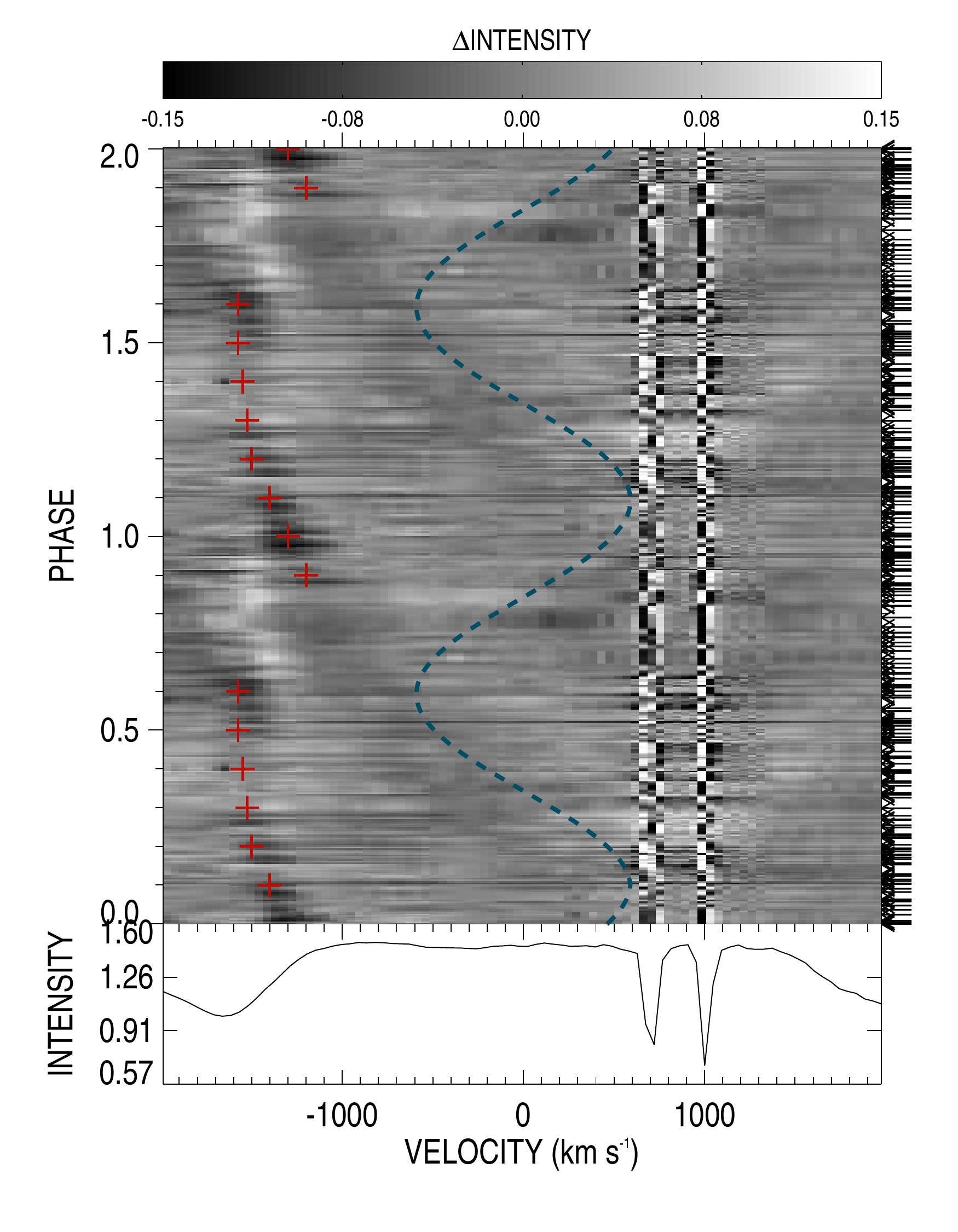}
\includegraphics[width=0.69\columnwidth]{HeIIgroup2.pdf}
\caption[The phased difference graycsale images of the C~{\sc IV}, He~{\sc I} and He~{\sc II} emission lines with the kinematics model.]{The phased difference grayscale images of the C~{\sc IV}~$\lambda\lambda$5802,5812 doublet, with a mean wavelength of $\lambda$5806 adopted ({\it left}), the He~{\sc I}~$\lambda$5876 ({\it center}) and He~{\sc II}~$\lambda$5411 ({\it right}) emission lines during the second 40-day period of the campaign. The DAC-like feature is outlined ({\it red crosses}) at negative velocities in the He~{\sc I}~$\lambda$5876 line. The dashed blue line represents the kinematics model calculated in Section 4.2.}
\label{fig:fig15}
\end{center}
\end{figure*}

While He~{\sc II}~$\lambda$5411 is a strong, isolated line, we also examined two other emission lines. Two phased grayscale difference images were created during the time period of HJD 2456478 - 2456518, when the central CIR was the strongest; one of the C~{\sc IV}~$\lambda\lambda$5802,5812 \AA\ doublet and the other of He~{\sc I}~$\lambda$5876 \AA. These are displayed in Fig.~\ref{fig:fig15}. As several spectrographs were unable to record these lines in parallel, these results required us to use a subset of the data to analyze the results.

\begin{figure}
\begin{center}
\includegraphics[width=0.75\columnwidth,angle=90]{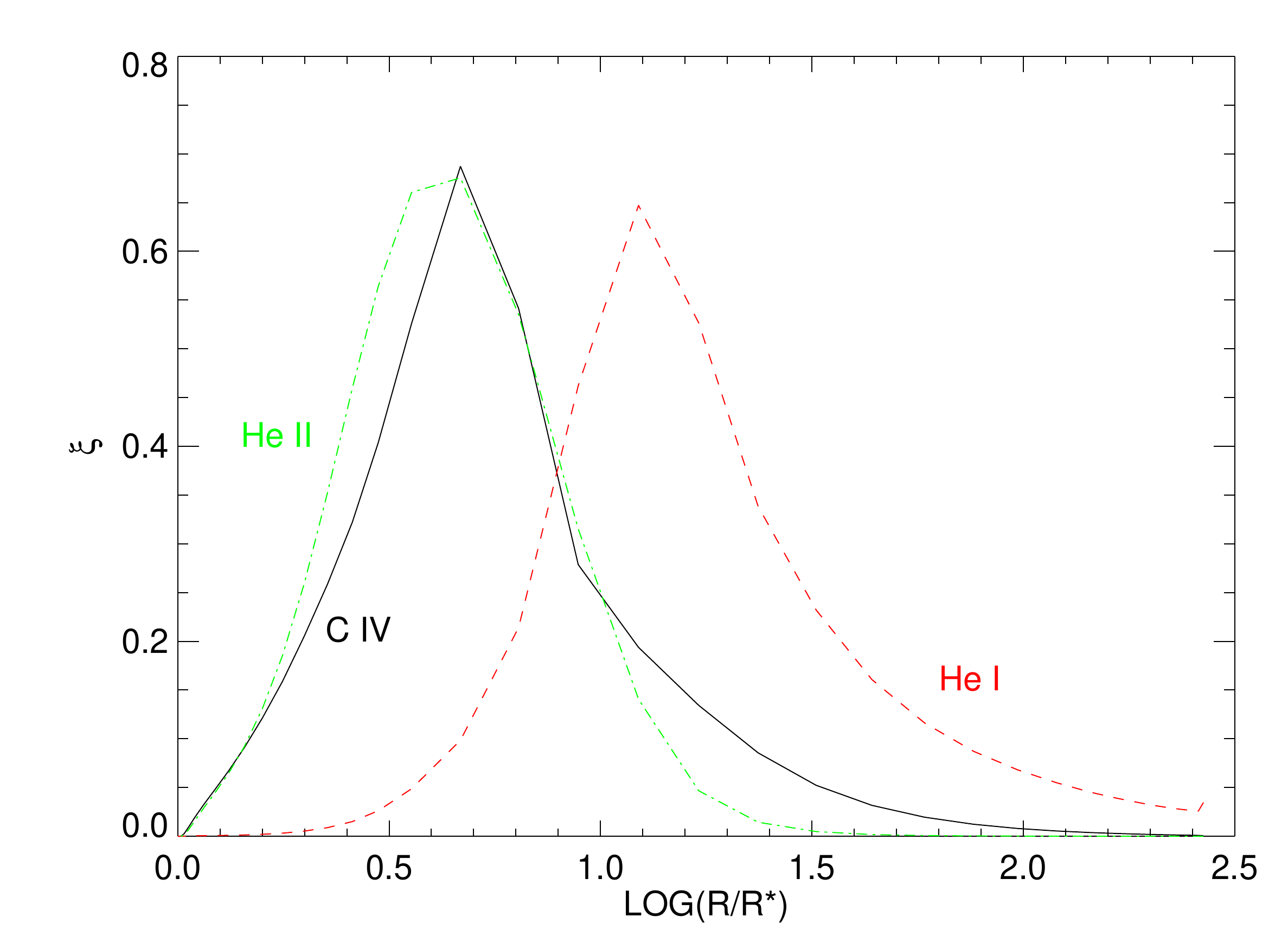}
\caption[The line-blanketing non-LTE model for WR 134 for the C~{\sc IV}, He~{\sc I} and He~{\sc II} formation zones.]{The emission measure $\xi (r)$ \citep{Hillier1987} as a function of the logarithmic radial distance of WR 134 using $v_{\infty} \sim$ 1900 km s$^{-1}$ for the C~{\sc IV}~$\lambda$5802, He~{\sc I}~$\lambda$5876 and He~{\sc II}~$\lambda$5411 formation zones. It can be seen that the He~{\sc I} formation zone is  the furthest out from the star of the three lines.}
\label{fig:fig17}
\end{center}
\end{figure}

The C~{\sc IV} emission line displays the same central CIR pattern that is detectable in the He~{\sc II} line, but this line is much fainter than He~{\sc II}~$\lambda$5411 and thus the data appear noisier and the CIR trace is more difficult to see on the grayscale images. For comparison purposes, we have superposed on the C~{\sc IV} grayscale image the fit of the He~{\sc II} central CIR trace. The agreement is very good. This is not surprising since they have similar ionization energies, with He~{\sc II} being 54.42 eV and  C~{\sc IV} being 64.49 eV, and are formed in very similar regions of the wind. This can be seen in Fig.~\ref{fig:fig17}, which displays the line formation zones of each of the three emission lines as a function of radial distance, using the Potsdam Wolf-Rayet (PoWR) line-blanketed, non-LTE model atmospheres found in \citet{Grafener2002} and \citet{Hamann2003} tailored for WR 134. The He~{\sc I} emission line also faintly displays the strong CIR pattern seen in He~{\sc II}, but in addition shows a DAC-like feature on the blue side of the line, moving from $v \simeq$ $-$1200 km s$^{-1}$ at $\phi =$ 0.90 to $v \simeq$ $-$1600 km s$^{-1}$ at $\phi =$ 1.65 (or $-$0.10 to 0.65). We have added red crosses to Fig.\ref{fig:fig15} to indicate the position of the DAC. The velocity of this absorption feature is compatible with that of the second incomplete CIR trace and therefore we associate the DAC to this CIR rather than with the central one. Furthermore, it appears around phase 0.85, which is approximately where the CIR in emission would appear if it were complete.

One interesting feature of the faint CIR pattern seen in emission in this transition is that it appears shifted in phase by $\sim$ 0.15 compared to the fit of the He~{\sc II} grayscale. This shift is most likely due to the line formation zone of He~{\sc I} being located further out in the wind than that of He~{\sc II} (see Fig.~\ref{fig:fig17}) and therefore the CIR feature will show a slight delay as it rotates. Such a shift in phase in the grayscale image for lines formed at different distances in the wind was in fact predicted by \citet{Dessart2002} in their modelling of CIRs. The C~{\sc IV} and He~{\sc I} lines show how the CIR feature within the C~{\sc IV} line affects the DAC features within the blue side of the He~{\sc I} line due to their close proximity. One can see that the absorption pattern of He~{\sc I}~$\lambda$5876 is partially filled in by the emission pattern of C~{\sc IV}~$\lambda\lambda$5802,5812 near phase 0.4. 

\subsection{Small-scale structures}

\begin{figure}
\begin{center}
\includegraphics[width=0.99\columnwidth]{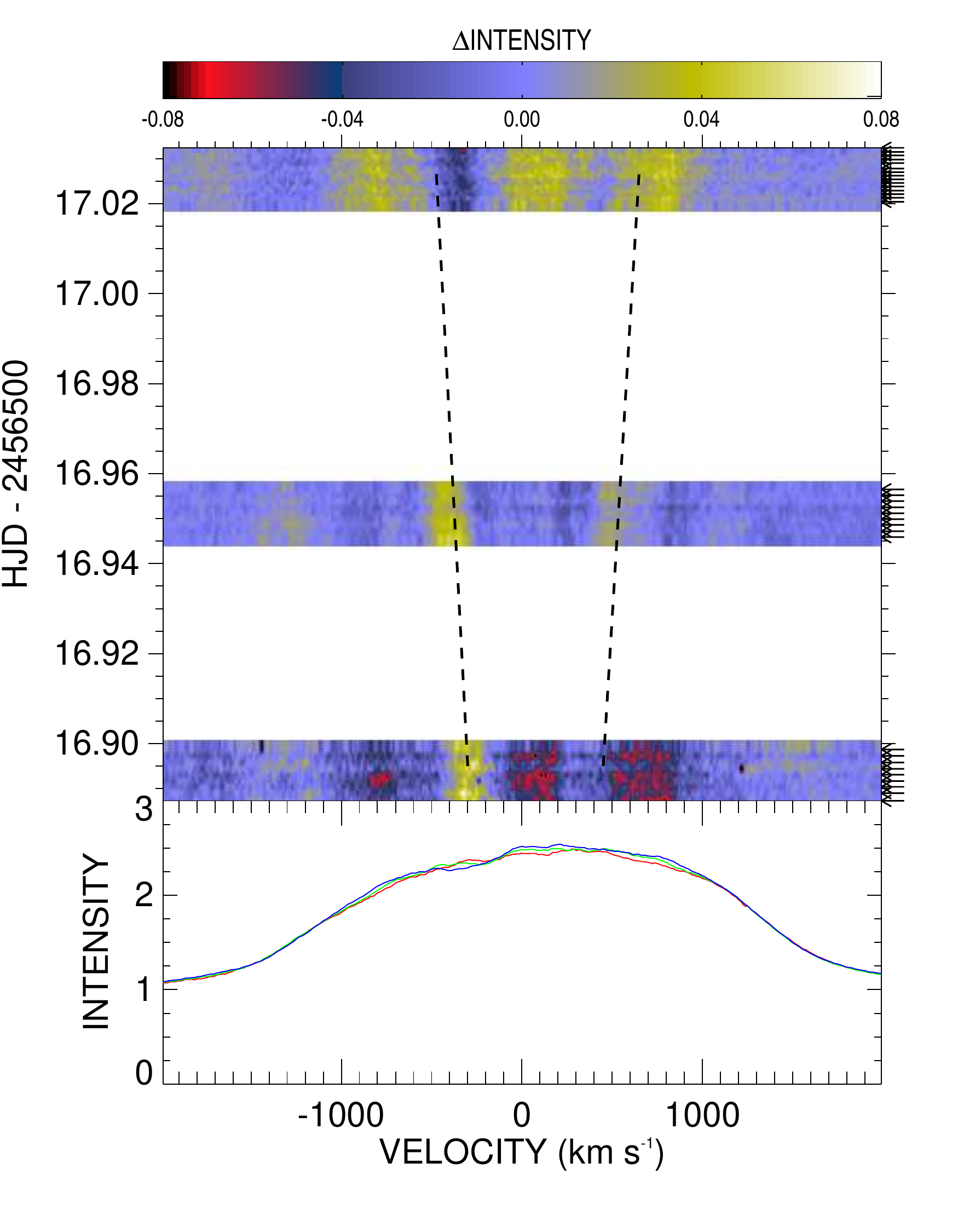}
\caption[Difference spectra from the mean of the Keck data displaying the movement of clumps over time.]{The difference spectra from the overall average of the He~{\sc II} $\lambda$5411 emission line of the Keck data. The bottom panel shows the average for each set ({\it red} first group, {\it green} second group and {\it blue} third group), with the top panel showing the intensity of each difference spectrum with respect to the mean Keck spectrum of the He~{\sc II}~$\lambda$5411 emission line in velocity space. The dashed lines indicate the movement of clumps over the night, while horizontal arrows indicate the exact position of each spectrum in time.}
\label{fig:fig19}
\end{center}
\end{figure}

WR 134 was first discovered to present small-scale structures within its wind by \citet{Moffat1988}, who also observed the He~{\sc II}~$\lambda$5411 emission line. They observed this star using the Canada-France-Hawaii Telescope (CFHT), along with another WN6 star, WR 136, and their data had an average S/N ratio of $\sim$ 300. Data were collected over one night for WR 134 and small structures were seen moving away from the line centre on timescales of hours, through analysis of the difference spectra. The star WR 136, however, did not show such features at the time, likely due to insufficient S/N. In order to search for these structures within our campaign data, two conditions had to be met: 1) the data must all be taken continuously with no large time gaps and 2) the S/N ratio and resolution must be high enough to rule out any possibility of noise hindering the detection. Because the Teide data had an order merge within the emission line, they became unreliable for the study of the small-scale structures as this increased the noise significantly. This left only the Keck data that met the conditions needed for this study, with a total of 35 spectra obtained using an echelle spectrograph (see Table~\ref{tab:tab1}) during a single night in three groups. Fig.~\ref{fig:fig19} displays the average of each of the 3 data groups for this line.

In order to identify variations caused by clumps, we must first understand the changes due to the CIRs within this observing night. From Fig.~\ref{fig:fig19}, the intensity change of the CIR excess emission can be clearly seen as a gradual increase of the flux during the night on the red-side of the line while the blue-side remains relatively constant. Difference spectra were calculated using the mean spectrum of the Keck data alone, which rendered the search for any short timescale changes in the emission line easier. The effect of the CIR is visible once more on the red side of the line by a gradual passage from mainly negative flux differences in group 1 to mainly positive flux differences in group 3. In addition, and in spite of the paucity of the data, a few clumps are detected on the red-side and blue-side of the central line and are found to move red-ward and blue-ward in time, respectively. These are displayed as dashed lines in Fig.~\ref{fig:fig19}.

Since WR 134 was already known to show clump structures within its wind \citep{Moffat1988}, one question that we aimed to answer during this campaign was whether these clumpy structures are always present and if they can be detected while the CIR has a strong presence within the wind. With this dataset, we have been able to determine that clumps and CIRs do indeed co-exist, although we were unable to study the details of the kinematics of the clumps with respect to that of the CIR due to the sparsity of our data within a given observing night.

\section{CONCLUSIONS AND DISCUSSION}

A 4-month professional/amateur campaign was organized for the summer of 2013 in order to observe the variations of three Wolf-Rayet stars. With six professional facilities and several amateurs, we were able to obtain a total of 392 spectra on WR 134 with almost continuous coverage during the campaign. The following conclusions have been found during the analysis of the spectroscopic data of WR 134:
\begin{itemize}
\item The  period of the variability was refined to 2.255 $\pm$ 0.008 (s.d.) days from the spectroscopic data around the He~{\sc II}~$\lambda$5411 emission line by calculating the moments of the line (Table~\ref{tab:tab1} and Appendix~B, online only) and calculating a \citet{Scargle1982} periodogram in velocity space (Fig.~\ref{fig:fig6}).
\item After a two-dimensional cross-correlation function (CCF) was applied to these difference plots, a $\sim$ 40-day time-coherency was seen after fitting a Gaussian function to the CCF (Fig.~\ref{fig:fig10}). A time-dependent \citet{Scargle1982} analysis of each moment also showed stronger intensities for a 40-day time interval near the centre of the campaign (Fig.~\ref{fig:fig5}), concluding that the CIR features have an overall lifetime of 40 $\pm$ 6 days, or $\sim$ 18 cycles.
\item We quantitatively compared the difference images to the CIR models created by \citet{Dessart2002}, leading us to conclude that it is an acceptable model to reproduce our observations and that the signature of two CIRs were in fact present within the grayscale image, and therefore in the wind of WR 134 during our observing campaign (Fig.~\ref{fig:fig14}). The emission traces corresponding to these two CIRs reached different maximum velocities, strongly hinting that they emerge from different latitudes on the surface of the star. Furthermore, these different maximum velocities were reached at different phases. The phase shift of $\Delta \phi \simeq$ 0.25 leads us to conclude that the CIRs are not found at the same longitude on the star but instead have a separation of $\Delta \phi \simeq$ 90$^{\circ}$.
\item After also analyzing the C~{\sc IV}~$\lambda$5806 and He~{\sc I}~$\lambda$5876 emission lines, we determined that the strong CIR is also detectable within the C~{\sc IV} line, while a DAC-like feature can be seen on the blue side of the He~{\sc I} line along with the central CIR trace, which for that line is slightly shifted in phase compared to what is seen in the other two lines (Fig.~\ref{fig:fig14}). This difference in phase is in qualitative agreement with the calculated models of \citet{Grafener2002} and \citet{Hamann2003}, showing the formation regions of the three emission lines analyzed (Fig.~\ref{fig:fig17}).
\item Small-scale features have been previously detected in this star \citep{Moffat1988} and were also studied using this spectroscopic dataset. Only a small portion of our observations was usable for this particular study, due to the difficulty in detecting clumps with a strong CIR present in the wind. The spectra provided by Keck Observatory had a high enough resolution and S/N ratio, with a total of 35 obtained in a single night. The difference plot created using the mean spectra of the Keck data (Fig.~\ref{fig:fig19}) shows smaller features in the wind, which allow us to conclude that these features do in fact coexist with the CIRs.
\end{itemize}

In addition to this data set detailing the variability of WR~134, spectroscopy was collected on WR~135 (WC8) and WR~137 (WC7pd+O9) during the campaign. WR~135 shows large amounts of variability from clumping \citep{Marchenko2006, Lepine2000} and WR~137 is a known dust-producing long-period binary \citep{Lamontagne1996, Williams2001} and has a potential CIR present in the WR wind \citep{Lefevre2005}. The results presented here highlight the importance of such spectroscopic time-series for understanding the variability and structure of WR winds. Indeed, a large number of treasures will be found in the analysis of the 2013 WR campaign data in the near-future. These results would not have been possible without the enthusiasm of the amateur astronomer community.

\section*{Acknowledgements}

The professional authors of this paper are grateful to the amateur astronomers of the Teide team, who invested personal time, money, and enthusiasm in this project. NSL and AFJM are grateful for financial aid from NSERC (Canada) and FQRNT (Qu\'ebec). NDR acknowledges postdoctoral support by the University of Toledo and by the Helen Luedtke Brooks Endowed Professorship. JHK acknowledges financial support to the DAGAL network from the People Programme (Marie Curie Actions) of the European Union's Seventh Framework Programme FP7/2007-2013/ under REA grant agreement number PITN-GA-2011-289313, and from the Spanish Ministry of Economy and Competitiveness (MINECO) under grant number AYA2013-41243-P. He also thanks the Astrophysics Research Institute of Liverpool John Moores University for their hospitality, and the Spanish Ministry of Education, Culture and Sports for financial support of his visit there, through grant number PR2015-00512. JK and BK are thankful for the support by the grant 13-10589S (GA \v{C}R). TR
acknowledges support from the Canadian Space Agency grant FAST.

We acknowledge the help and support of collegues at the Universit\'e de Montr\'eal, W. M. Keck Observatory, Universit\"at Potsdam, Ond\v{r}ejov Observatory (Perek 2-m Telescope) and Teide Observatory. The 0.82m IAC80 Telescope is operated on the island of Tenerife by the Instituto de Astrof\'isica de Canarias (IAC) in the Spanish Observatorio del Teide. Based on observations made with the Nordic Optical Telescope, operated by the Nordic Optical Telescope Scientific Association at the Observatorio del Roque de los Muchachos, La Palma, Spain, of the IAC. We also acknowledge the support from the amateur spectroscopy groups, VdS and ARAS. The authors wish to recognize and acknowledge the very significant cultural role and reverence that the summit of Mauna Kea has always had within the indigenous Hawaiian community. We are most fortunate to have the opportunity to conduct observations from this mountain. 




\bibliographystyle{mnras}
\bibliography{references} 





\bsp	
\label{lastpage}
\end{document}